\pdfoutput=1 
\documentclass[aps,prd,twocolumn,superscriptaddress,tightelines,nofootinbib]{revtex4}
\usepackage{graphicx}
\usepackage{rotating}
\usepackage{epsfig}
\usepackage{bm}
\usepackage{latexsym,amssymb,amsmath,amsfonts,amssymb,txfonts,pxfonts,wasysym,float}
\usepackage{multirow}
\usepackage{color}
\newcommand{\beq}[1]{\begin{equation}\label{#1}}
\newcommand{\eeq}{\end{equation}}
\newcommand{\bea}[1]{\begin{eqnarray} \label{#1}}
\newcommand{\eea}{\end{eqnarray}}
\newcommand{\ba}{\begin{array}}
\newcommand{\ea}{\end{array}}

\def\be{\begin{equation}}
\def\ee{\end{equation}}
\def\gs{\mathrel{
   \rlap{\raise 0.511ex \hbox{$>$}}{\lower 0.511ex \hbox{$\sim$}}}}
\def\ls{\mathrel{
   \rlap{\raise 0.511ex \hbox{$<$}}{\lower 0.511ex \hbox{$\sim$}}}}

\newcommand{\postscript}[2]{\setlength{\epsfxsize}{#2\hsize}
   \centerline{\epsfbox{#1}}}
\newcommand{\e}{\varepsilon}

\newcommand{\comment}[1]{}

\usepackage[usenames,dvipsnames]{xcolor}
\definecolor{orange}{cmyk}{0,0.5,1,0}
\definecolor{rossoCP3}{cmyk}{0,.88,.77,.40}
\definecolor{graa}{rgb}{0.8,0.8,0.8}
\definecolor{blaa}{rgb}{0.2,0.2,0.6}

\begin{document}

\title{\color{rossoCP3}{Ultrahigh-Energy Cosmic Ray  Composition 
    from
    the Distribution  of Arrival Directions 
}}

\author{Rita C. dos Anjos}
\affiliation{Department of Physics \& Astronomy,  Lehman College, City University of
  New York, NY 10468, USA}
\affiliation{Departamento de Engenharias e Exatas, Universidade Federal do Paran\'a,  85950-000 Palotina, Brazil}

\author{Jorge F. Soriano}
\affiliation{Department of Physics \& Astronomy,  Lehman College, City University of
  New York, NY 10468, USA}
\affiliation{Department of Physics,
 Graduate Center, City University
  of New York,  NY 10016, USA}

\author{Luis A. Anchordoqui}
\affiliation{Department of Physics \& Astronomy,  Lehman College, City University of
  New York, NY 10468, USA}
\affiliation{Department of Physics,
 Graduate Center, City University
  of New York,  NY 10016, USA}
\affiliation{Department of Astrophysics,
 American Museum of Natural History, NY
 10024, USA}

\author{Thomas \nolinebreak C. \nolinebreak Paul}
\affiliation{Department of Physics \& Astronomy,  Lehman College, City University of
  New York, NY 10468, USA}

\author{Diego F. Torres}

\affiliation{Institute of Space Sciences (ICE-CSIC),  Campus UAB,
  Carrer de Magrans s/n, 08193 Barcelona, Spain 
}
\affiliation{Instituci\'o Catalana de Recerca i Estudis Avan\c{c}ats
  (ICREA),  E-08010 Barcelona, Spain
}

\affiliation{Institut d'Estudis Espacials de Catalunya (IEEC),
08034 Barcelona, Spain
}

\author{John F. Krizmanic}
\affiliation{NASA Goddard Space Flight Center, Greenbelt, MD, USA}

\author{Timothy A. D. Paglione}
\affiliation{Department of Physics,
 Graduate Center, City University
  of New York,  NY 10016, USA}
\affiliation{Department of Astrophysics,
 American Museum of Natural History, NY
 10024, USA}

\affiliation{Department of Earth \& Physical Sciences, York College,
  City University of New York,  NY 11451, USA}

\author{Roberto \nolinebreak J. \nolinebreak Moncada}

\affiliation{Department of Astrophysics,
 American Museum of Natural History, NY
 10024, USA}

\affiliation{Department of Physics, University of Miami, Coral Gables,
  FL  33146, USA}

\author{Frederic Sarazin}
\affiliation{Department of Physics, Colorado School of Mines, Golden, CO 80401, USA}

\author{Lawrence Wiencke}
\affiliation{Department of Physics, Colorado School of Mines, Golden, CO 80401, USA}

\author{Angela V. Olinto}
\affiliation{Department of Astronomy \& Astrophysics, University of Chicago, Chicago, IL 60637, USA}
\affiliation{
Enrico Fermi Institute and Kavli Institute for Cosmological Physics,
University of Chicago, Chicago, IL 60637, USA}

\begin{abstract}
  \noindent The sources of ultrahigh-energy cosmic rays (UHECRs) have
  been difficult to catch. It was recently pointed out that while
  sources of UHECR protons exhibit anisotropy patterns that become
  denser and compressed with rising energy, nucleus-emitting-sources
  give rise to a {\it cepa stratis} (onion-like) structure with layers that become
  more distant from the source position with rising energy. The
  peculiar shape of the hot spots from nucleus-accelerators is steered
  by the competition between energy loss during propagation and
  deflection on the Galactic magnetic field (GMF). Here, we run a full-blown
  simulation study to accurately characterize the deflections of UHECR
  nuclei in the GMF.  We show that while the {\it cepa stratis}
  structure provides a global description of anisotropy patterns
  produced by UHECR nuclei {\it en route} to Earth, the hot spots are
  elongated depending on their location in the sky due to the regular
  structure of the GMF. We demonstrate that with a high-statistics
  sample at the high-energy-end of the spectrum, like the one to be
  collected by NASA's POEMMA mission, the energy dependence of the
  hot-spot contours could become a useful observable to identify the
  nuclear composition of UHECRs. This new method to determine the
  nature of the particle species is complementary to those using
  observables of extensive air showers, and therefore is unaffected by
  the large systematic uncertainties of hadronic interaction models.
\end{abstract}

\maketitle

\section{Introduction}

The search for the sources of ultrahigh-energy cosmic rays (UHECRs)
continues to be one of the most challenging and at the same time most
important tasks in astrophysics~\cite{Anchordoqui:2018qom,Kotera:2011cp}. UHECR deflection by intervening
magnetic fields hampers pinning down their origins.  The
most recent data (interpreted using various hadronic models) seem to
indicate that the most energetic $(E \agt 10^{10}~{\rm GeV})$ cosmic
rays may not just be protons, but rather atomic nuclei of charge
$Ze$~\cite{Aab:2014kda,Aab:2014aea,Aab:2016htd,Abbasi:2015xga}. Even
though cosmic ray trajectories would naturally undergo less magnetic
bending as the kinetic energy (rigidity) is increased, UHECR nuclei
are expected to suffer significant deflections while traversing the
Galaxy.

To makes things worse, magnetic fields are
not well constrained by current data. If we endorse recent models of
the Galactic magnetic
field (GMF)~\cite{Pshirkov:2011um,Jansson:2012pc,Jansson:2012rt,Unger:2017kfh},
then typical values of the deflections of UHECRs crossing the Galaxy
are
\begin{equation}
\theta \sim 10^\circ \ Z \
\left(\frac{E}{10^{10}~{\rm GeV}}\right)^{-1} \,,
\label{deflection}
\end{equation}
depending on the direction considered~\cite{Erdmann:2016vle,Farrar:2017lhm}. Therefore, tracing the origin
of a particular UHECR nucleus back to its source in the sky is
not trivial.

\begin{figure}[tpb]
\postscript{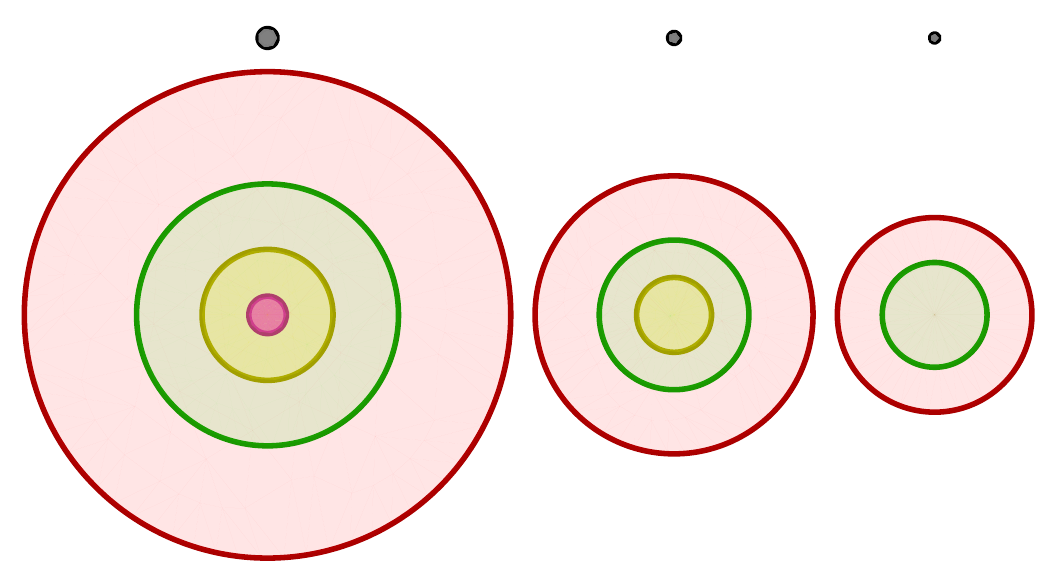}{0.99}
\begin{center}
\caption{Circles representing the composition-layered structure of hotspots at different energies, for proton sources (top) and nuclei sources (bottom). The radii of the circles respect the proportions of the angular sizes given by (\ref{deflection}), for protons (black), helium (magenta), nitrogen (yellow), silicon (green) and iron (red); and for $40\,\mathrm{EeV}$ (left), $70\,\mathrm{EeV}$ (center) and $100\,\mathrm{EeV}$ (right).
\label{fig:0}}
\end{center}
\end{figure}

UHECR nuclei lose energy
{\it en route} to Earth through interactions with the universal
radiation
fields~\cite{Greisen:1966jv,Zatsepin:1966jv,Puget:1976nz}. Thus and
so, it was recently pointed out that the combination of magnetic
deflections and energy losses during propagation should produce an
unequivocal anisotropy pattern for accelerators of UHECR nuclei: a
{\it cepa stratis} structure with layers that increase with rising
energy~\cite{Anchordoqui:2017abg}.  This is in sharp contrast to
anisotropy patterns of pure-proton sources, which become denser and
compressed with rising energy. The combination of these effects leads
to an onion-like layered structure depending on composition and
energy. To visualize this, first recall that if the sources emitted
only protons, the size of the corresponding ``spot'' should decrease
with rising energy due to reduced deflection in magnetic fields. In contrast,
if sources produce a mixed composition, a different quality
emerges. Lighter compositions tend to shorter mean-free-paths at
higher energies, so as their energy increases they
begin to disappear from the sample leaving behind only the lower
energy component. The latter suffers a relatively smaller magnetic deflection compared to heavier nuclei at all energies. One thus ends up with a {\it cepa stratis} structure in
which the energies of the species observed closer to the source have a
{\em lower} rather than higher energy, as they would in the case that
the sources emitted only protons. This effect is shown schematically
in Fig.~\ref{fig:0}.

In this paper we simulate realistic UHECR sky maps for a wide range of
possible nuclear species and study individual anisotropy patterns from
nearby sources to quantify the variation in (shape and size) of the
expected ``squeezed onion layers.''  We also present a statistical
test to isolate the UHECR nuclear composition using a subsample of
the distribution of arrival directions associated with a particular
source in the cosmic-ray-sky. All source types are represented within
our cosmic backyard for light and heavy nuclei, so all source types
are {\it a priori} candidates for the nearby exploration.

The layout of the paper is as follows. We begin in Sec.~\ref{2ish}
with an overview of the main characteristics of potential nearby
sources. In Sec.~\ref{sec2} we study the energy losses during
propagation and in Sec.~\ref{sec3} the deflections on the GMF. Armed
with our findings, in Sec.~\ref{sec4} we develop a statistical test to
probe the nuclear composition of UHECR using the distribution of
arrival directions. After that, in Sec.~\ref{sec5} we demonstrate that
NASA's Probe Of Extreme Multi-Messenger Astrophysics (POEMMA)
mission~\cite{Olinto:2017xbi} will attain sensitivity to clarify the
nuclear composition of recently reported hot spots by the Telescope
Array and Pierre Auger collaborations~\cite{Biteau:2018paris}. The
paper wraps up with some conclusions presented in Sec.~\ref{sec6}.

\section{Experimental data}
\label{2ish}

\subsection{UHECR Anisotropies}
Over the years, stronger and stronger experimental evidence has been
accumulating indicating a possible correlation between the arrival
directions of the highest energy cosmic rays and nearby starburst
galaxies~\cite{Anchordoqui:2002dj,Aab:2018chp}. Recently, using data
collected by the Pierre Auger Observatory, the hypothesis of UHECR
emission from the 23 brightest nearby starburst galaxies (SBGs) with a
radio flux larger that 0.3~Jy (selected out the 63 objects within
250~Mpc search for gamma-ray emission by the Fermi-LAT
Collaboration~\cite{Ackermann:2012vca}) was tested against the null
hypothesis of isotropy through an unbinned maximum-likelihood
analysis~\cite{Aab:2018chp}. The adopted test statistic (TS) for
deviation from isotropy being the standard likelihood ratio test
between the starburst-generated UHECR sky model and the null
hypothesis.  The TS was maximized as a function of two free parameters
(the angular radius common to all sources, which accounts in an
effective way for the magnetic deflections, and the signal fraction),
with the energy threshold varying in the range $10^{10.3} \alt E/{\rm
  GeV} \alt 10^{10.9}$. For a given energy threshold, the TS for
isotropy follows a $\chi^2$ distribution with two degrees of
freedom. The TS is maximum above $10^{10.6}~{\rm GeV}$, with a local
$p$-value of $3 \times 10^{-6}$. The smearing angle and the
anisotropic fraction corresponding to the best-fit parameters are
${13^{+4}_{-3}}^\circ$ and $(10 \pm 4)\%$, respectively. Remarkably,
the energy threshold of largest statistical significance coincides
with the observed suppression in the
spectrum~\cite{Abraham:2008ru,Abraham:2010mj,Aab:2017njo}, implying
that when we properly account for the barriers to UHECR propagation in
the form of energy loss
mechanisms~\cite{Greisen:1966jv,Zatsepin:1966jv} we obtain a self
consistent picture for the observed UHECR horizon. The scan in energy
thresholds comes out with a penalty factor, which was estimated
through Monte-Carlo simulations.  The post-trial chance probability in
an isotropic cosmic ray sky is $4.2 \times 10^{-5}$, corresponding to
a 1-sided Gaussian significance of $4\sigma$~\cite{Aab:2018chp}.

Auger data ($E \agt 10^{10.77}~{\rm GeV}$) also show a slightly weaker
association ($2.7\sigma$) with active galactic nuclei (AGNs) that emit
$\gamma$-rays (a.k.a. $\gamma$AGNs) from the 2nd catalogue of hard
{\it Fermi}-LAT sources (2FHL)~\cite{Ackermann:2015uya}.  The maximum
deviation for $\gamma$AGNs is found at an intermediate angular scale
of ${7^{+4}_{-2}}^\circ$ with an anisotropic fraction of $(7 \pm
4)\%$~\cite{Aab:2018chp}.

On a separate track, the Telescope Array (TA) has recorded a
statistically significant excess in cosmic rays, with energies above
$10^{10.75}~{\rm GeV}$, above the isotropic background-only
expectation~\cite{Abbasi:2014lda,Kawata:2015whq}. This is colloquially
referred to as the ``TA hot spot.'' The excess is centered at Galactic
coordinates $(l,b) \simeq (177^\circ,50^\circ)$, spanning a region of
the sky with $\sim 20^\circ$ radius. The chance probability of this
hot spot in an isotropic cosmic ray sky was calculated to be $3.7 \times
10^{-4}$, corresponding to 3.4$\sigma$.

\begin{figure*}[tpb]
\postscript{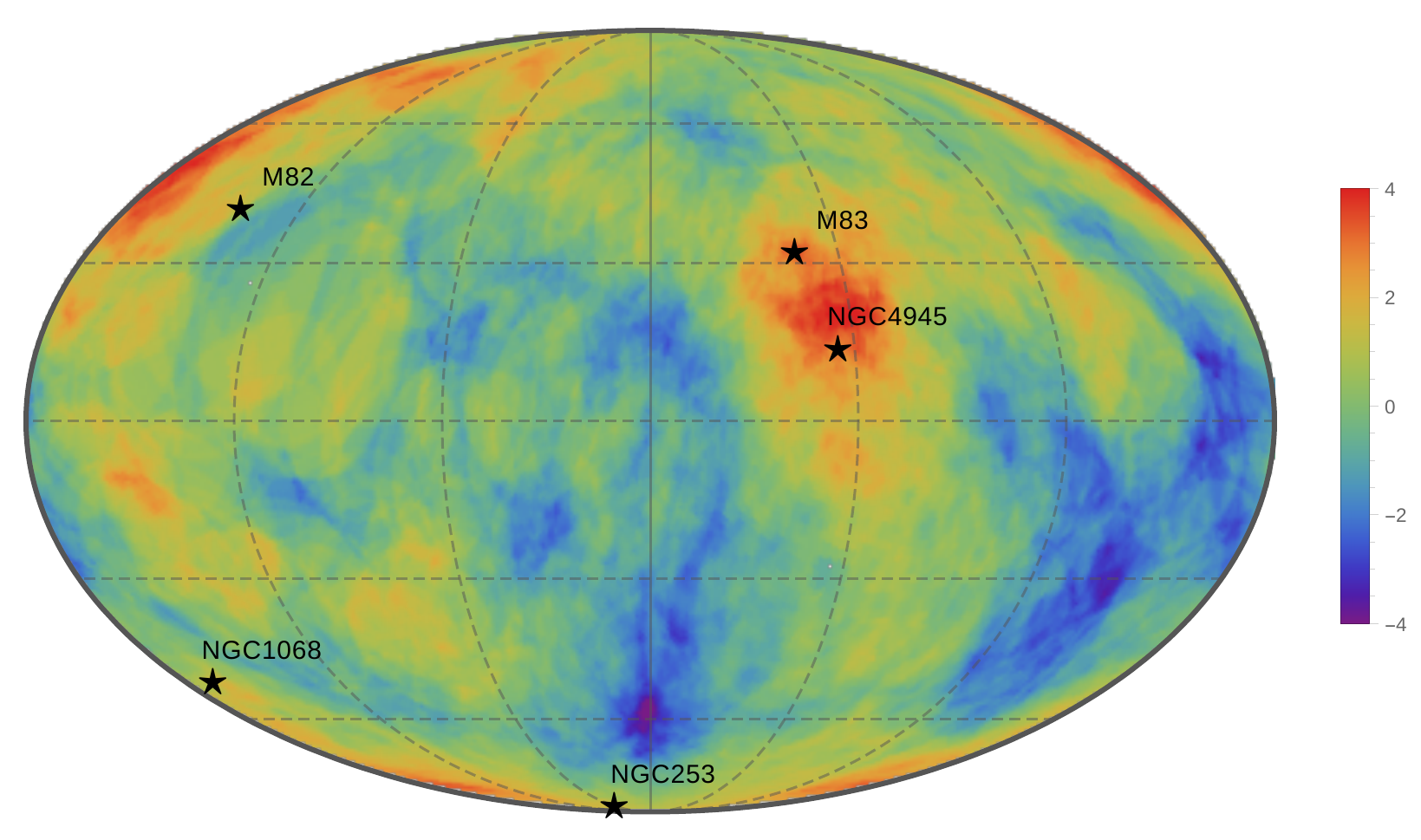}{0.99}
\begin{center}
\caption{Skymap in Galactic coordinates of the Li-Ma significances of
  overdensities in $20^\circ$ radius windows for  840 events recorded by
Auger with $E> E_{\rm Auger}$ and 130 events
recorded by TA with $E> E_{\rm TA}$~\cite{note}. The color scale
indicates the significance in units of standard deviations;
negative values follow the convention of indicating the (positive)
significance of deficits.
\label{fig:0.5}}
\end{center}
\end{figure*}

The most recent search for hot spot anisotropies is a joint effort by
the two collaborations considering 840 events recorded by
Auger with $E> E_{\rm Auger} = 10^{10.6}~{\rm GeV}$ and 130 events
recorded by TA with $E > E_{\rm TA} = 10^{10.73}~{\rm
  GeV}$~\cite{Biteau:2018paris}. Before proceeding, we pause to note
that though the techniques for assigning energies to events are nearly
the same in both experiments, there are differences as to how the
primary energies are derived at Auger and TA, with systematic
uncertainties in the energy scale of the experiments amounting to
about $14\%$ and $21\%$ respectively, corresponding to about $70\%$
uncertainty in the flux above a fixed energy threshold. By comparison,
the uncertainties on the respective exposures are minor ($\alt 1\%$
and $\simeq 3\%$, respectively). Therefore, it is necessary to
cross-calibrate the energy scales of the two datasets to avoid
introducing a spurious North/South asymmetry due to an energy scale
mismatch. This is accomplished by exploiting the wide declination band
($- 16^\circ \alt \delta \alt +45^\circ$) where the two datasets
overlap. Regardless of the true arrival direction distribution, within
a region of the sky $\Delta \Omega$ fully contained in the field of
view (FoV) of both observatories, the sum over observed events $\sum_i
1/\omega({\bf n}_i )$ (where $\omega$ is the directional exposure of
each observatory in the direction ${\bf n}_i$, in ${\rm km \, yr}$
units) is an unbiased estimator of $\int_{\Delta \Omega} \Phi({\bf n})
\, d {\bf n}$ (where $\Phi$ is the directional UHECR flux integrated
above the considered energy threshold, in ${\rm km^{-2} \, yr^{-1} \,
  sr^{-1}}$ units) and should be the same for both experiments except
for statistical fluctuations. This criterium is generally adopted to
cross-calibrate the energy scales and to determine $E_{\rm Auger}$ and
$E_{\rm TA}$ such that the Auger flux above $E_{\rm Auger}$ matches
the TA flux above $E_{\rm
  TA}$~\cite{diMatteo:2018vmr}.\footnote{Actually, the region of the
  sky which is mostly used spans the declination band $-12^\circ \leq
  \delta \leq +42^\circ$. This is because including directions too
  close to the edge of the FoV of one of the observatories would
  result in larger statistical fluctuations due to very large values
  of $1/\omega(\bf{n}_i)$ near the edge.} The most significant
excesses observed in a $20^\circ$ search are at Galactic longitude  and
latitude: $(l, b) \approx (303.0^\circ, 12.9^\circ)$ and
($l, b) \approx (162.5^\circ,44.4^\circ)$, with local
(Li-Ma~\cite{Li:1983fv}) statistical significance for the rejection of
the null (background only) hypothesis of $4.7\sigma$ and $4.2\sigma$,
respectively~\cite{Biteau:2018paris}. The Li-Ma significance map of
this data-sample is shown in Fig.~\ref{fig:0.5}.  The most significant
hot spot is near the location of starburst galaxies NGC 4945 and
M83. The possible association of the TA hot spot with M82 has not gone
unnoticed~\cite{Anchordoqui:2014yva,Fang:2014uja,He:2014mqa,Pfeffer:2015idq,Attallah:2018euc}. A
warm spot is also visible in the vicinity of NGC 253 near the Galactic
south pole.

Very recently, the TA Collaboration carried out a test of the reported
correlation between the arrival directions of UHECRs and
SBGs~\cite{Aab:2018chp}. The data sample for this analysis includes
cosmic rays with $E >E'_{\rm TA} = 43~{\rm EeV}$ detected by TA in a
nine year period from May 2008 to May 2017. These data are compatible
with isotropy to within $1.1\sigma$ and with Auger result to within
$1.4\sigma$, and so the TA Collaboration concluded that with their
current statistics they cannot make a statistically significant
corroboration or refutation of the reported possible correlation
between UHECRs and SBGs~\cite{Abbasi:2018tqo}. It is important to
note, however, that $E'_{\rm TA} < E_{\rm TA}$. Most importantly,
 $E_{\rm TA}$ is above the energy at which TA observes the
suppression in the spectrum~\cite{Abbasi:2007sv,AbuZayyad:2012ru}, but
$E'_{\rm TA}$ is below. This implies that
the data sample of the test carried out by the TA Collaboration is
most likely contaminated from the isotropic background of UHECRs emitted by far
away sources, and consequently this would tend to reduce the
significance of any possible correlation with nearby sources.

\begin{figure*}[tbp]
\begin{minipage}[t]{0.49\textwidth}
    \postscript{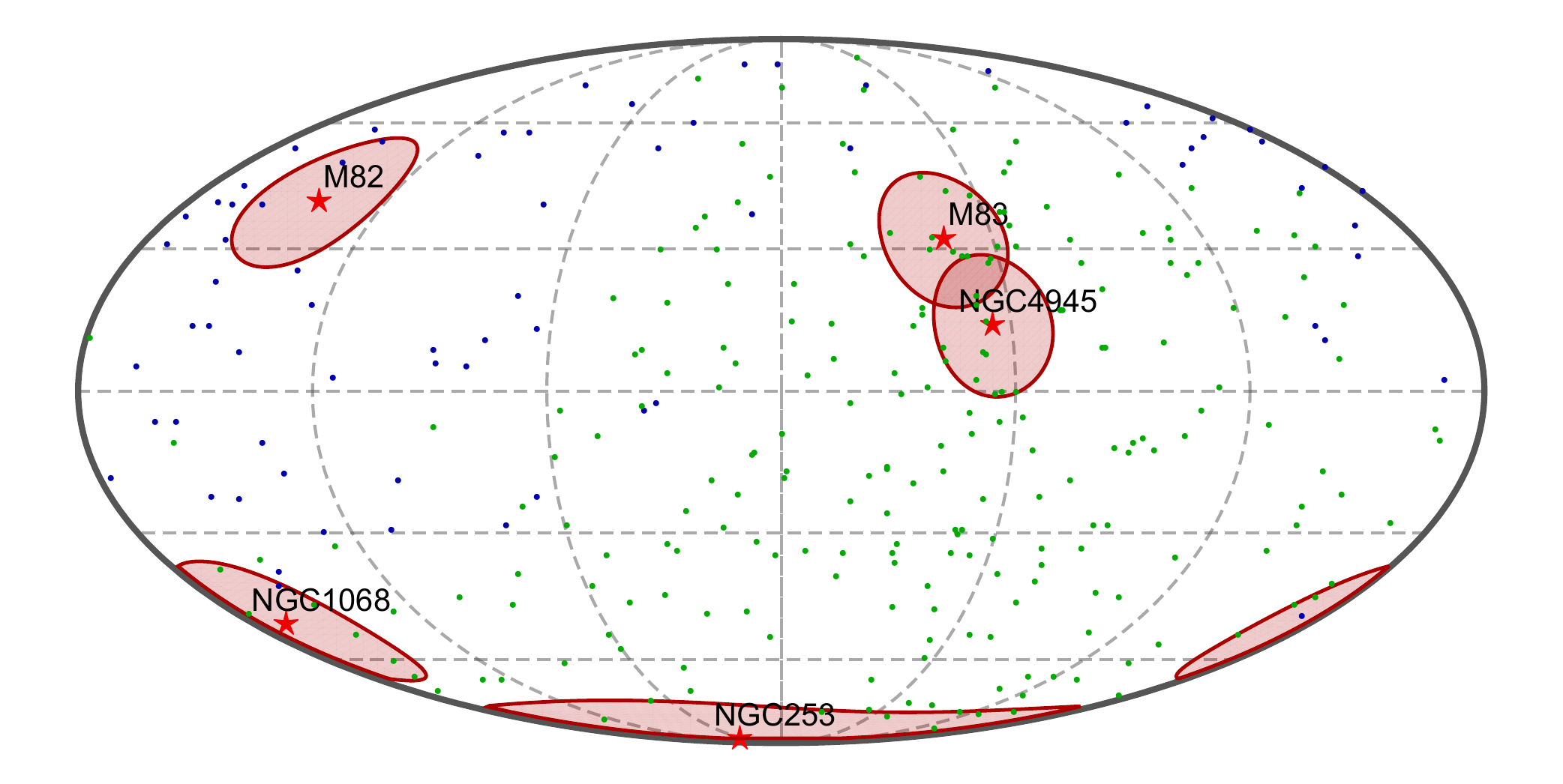}{0.9}
\end{minipage}
\hfill \begin{minipage}[t]{0.49\textwidth}
  \postscript{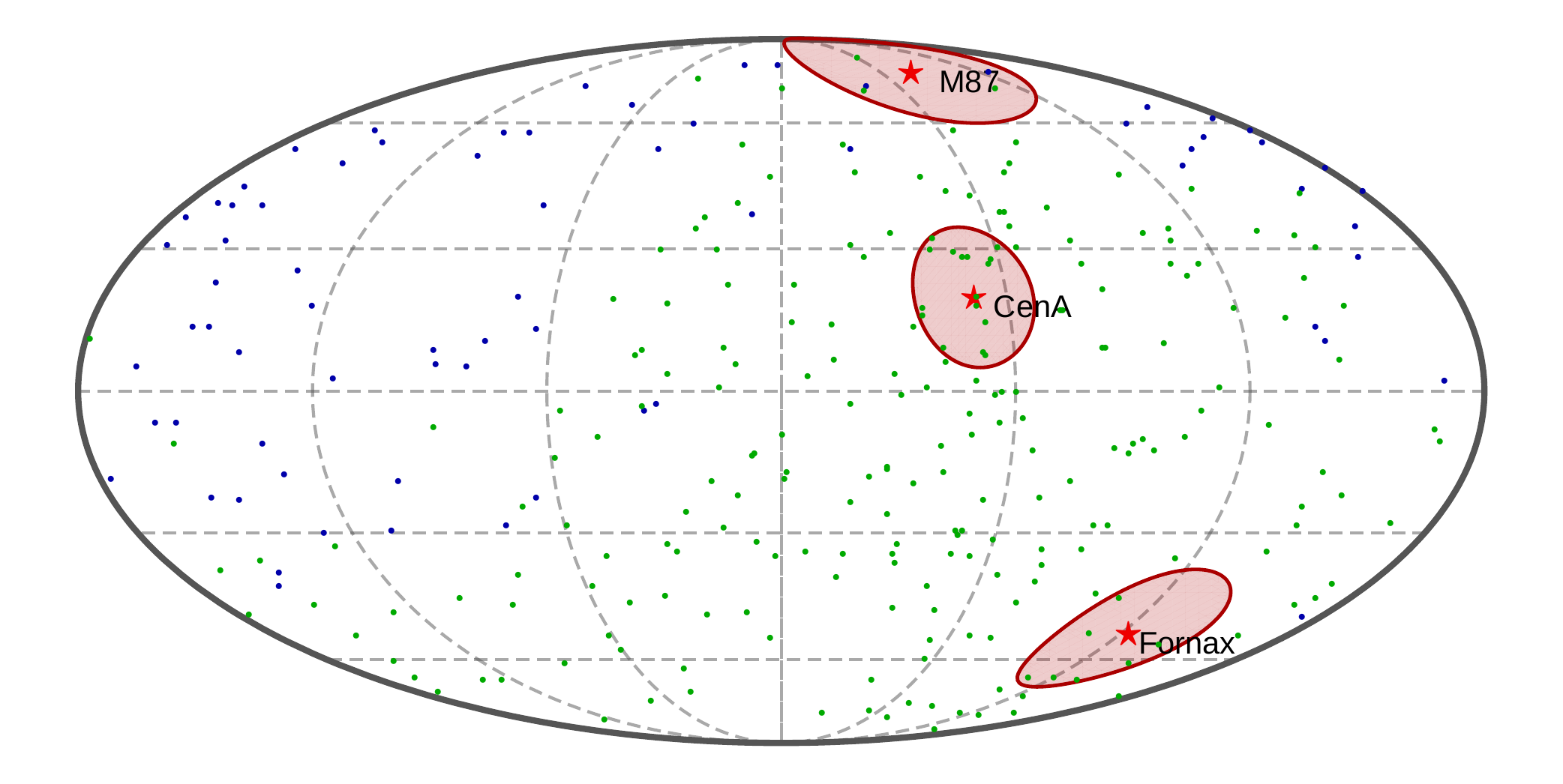}{0.9}
\end{minipage}
\caption{Comparison of UHECR event locations with starburst- (left)
  and radio-galaxies (right) in Galactic coordinates. The green points
  indicate the arrival directions of 231 events with $E > 52~{\rm
    EeV}$ and zenith angle $\theta < 80^\circ$ detected by the Pierre Auger
  Observatory from 2004 January 1 up to 2014 March 31~\cite{PierreAuger:2014yba}. The
  blue points indicate the arrival directions of 72 events with $E >
  57~{\rm EeV}$
  and $\theta < 55^\circ$ recorded from 2008 May 11 to 2013 May 4 with TA [22].
  The stars indicate the location of nearby starburst- (left) and
  radio-galaxies (right).
  The shaded regions delimit
 angular windows around the sources of angular radius of $15^\circ$.}
\label{fig:extra}
\end{figure*}

\subsection{Source Spectra}

It has long been suspected that the powerful jets and the mammoth
radio-lobes of nearby
$\gamma$AGNs~\cite{Biermann:1987ep,Rachen:1992pg} as well as the
galactic-scale superwinds of SBGs~\cite{Anchordoqui:1999cu} provide
profitable arenas for the formation of collisionless plasma shock
waves, in which UHECRs can be accelerated by bouncing back and forth
across the shock. In addition, because of the high prevalence of
supernovae, SBGs are thought to contain a large density of newly-born
pulsars, which can accelerate UHECRs via unipolar
induction~\cite{Blasi:2000xm,Arons:2002yj}.
 
Arguably, when all of the above is combined $\gamma$AGNs and SBGs
become the leading candidate sources at the very high energy end of
the spectrum. Therefore, we will adopt these astrophysical
objects as our working example.

\begin{table}
  \caption{Spectral indices of selected nearby sources: $\gamma$ maximizes the likelihood and $[\gamma_l,\gamma_r]$ indicates to a 68\% confidence interval of the
    spectral index.} 
\begin{tabular}{cccccc}
\hline
\hline
Source   & Dataset &   Events &    $\gamma$  &  $\gamma_l$ &    $\gamma_r$ \\
\hline
~~~~NGC 253~~~~ & ~~~~Auger~~~~	&	~~~~8~~~~	&	~~~~4.8~~~~	&	~~~~3.6~~~~	&	~~~~6.4~~~~\\
NGC 4945	   & Auger	&	14	&	6.8	&	5.4	&	8.5\\
M83		 &  Auger	&	13	&	4.6	&	3.7	&	5.7\\
NGC 1068	 & Auger	&	8	&	4.9	&	3.7	&	6.4\\
NGC 1068	 & TA	&	2	&	3.9	&	2.3   &	6.5\\
M82		& TA	&	3	&	5.3	&     3.3    &	8.3 \\
Cen A  &  Auger  &   16 &    5.5  &  4.5 &    6.8 \\            
Fornax A &   Auger &     7  &   7.0 &   5.0 &   9.5\\ 
M87  &  Auger   & 3  &  15.2 &   8.5 &   25.0 \\    
M87  &  TA  &  2 &   8.7 &   4.5 &   15.5 \\
\hline
\hline
\end{tabular}
\label{tabla_gamma}
\end{table}    

In order to describe the sources properly, we study the spectra of
Auger and TA events around selected objects that are relevant for this
analysis. We consider the data published
in~\cite{PierreAuger:2014yba,Abbasi:2014lda}, consisting on 231 events
above $52~{\rm EeV}$ detected by the Pierre Auger Observatory, and 72
events above $57~{\rm EeV}$ detected by TA. We select several sources
from Auger and TA searches of anisotropy. For each of those sources,
we define an angular window around their directions on the sky with
angular radius of $15^\circ$, as shown in Fig.~\ref{fig:extra}. This
value serves just as an orientation, and we do not imply that the
events from those sources should be contained in those angular
windows. Nevertheless, the analysis presented
in~\cite{PierreAuger:2014yba} results in such angular size for one of
the sources. We perform a maximum likelihood estimation of the
spectral index around each of the sources, for each of the data
samples (if there is more than one event), assuming a single power law
spectrum, $d N/d E\propto E^{-\gamma}$.  In Table~\ref{tabla_gamma} we
show the values of $\gamma$ maximizing the likelihood, as well as the
$68\%$ confidence level intervals $[\gamma_l,\gamma_r]$. All the
individual spectra are very steep, reflecting the suppression in the
nearly isotropic UHECR spectrum.

\subsection{Starburst Energetics}
It was recently pointed out that 
starburst superwinds struggle to meet the power requirements to
accelerate cosmic rays to the maximum observed energies~\cite{Matthews:2018laz}.  In detail,
the magnetic field $B$ carries with it an energy density $B^2/(8\pi)$
and the flow carries with it an energy flux $> u B^2/(8\pi)$, where
$u$ is the shock velocity. Thus, for an accelerator of size $R$, this sets a
lower limit on the rate at which the energy is carried by the out-flowing plasma,
\begin{equation}
L_B > \frac{1}{8} \ u \ R^2 \ B^2 ,
\label{eq:Waxman}
\end{equation}
and which
must be provided by the source~\cite{Waxman:1995vg}. Inserting typical parameters of SBGs ($L_B
\sim  10^{42.5}~{\rm erg/s}$, $R \sim 8~{\rm kpc}$, and $u \sim 
10^{3.3}~{\rm km/s}$~\cite{Anchordoqui:2018vji}) into (\ref{eq:Waxman}) leads to the constraint   $B< 15\,\mu{\rm G}$, and
consequently a Hillas maximum rigidity
\begin{equation}
{\cal R} \simeq (u/c) \ B \ R  <  10^{8.9} \,{\rm GV} .
\label{eq:Hillas}
\end{equation}
However, radio continuum and polarization observations of M82 provide
an estimate of the magnetic field strength in the core region of
$98~\mu{\rm G}$ and in the halo of $24~\mu{\rm G}$; see e.g. the
equipartition $B$ map in Fig.~16 of~\cite{Adebahr:2012ce}. Averaging
the magnetic field strength over the whole galaxy results in a mean
equipartition field strength of $35~\mu{\rm G}$. Independent magnetic
field estimates from polarized intensities and rotation measures yield
similar strengths~\cite{Adebahr:2017}. Comparable field strengths have
been estimated for NGC
253~\cite{Beck,Heesen:2008cs,Heesen:2009sg,Heesen:2011kj} and other
starbursts~\cite{Krause:2014iza}. Actually, the field strengths could
be higher if the cosmic rays are not in equipartition with the
magnetic field~\cite{Thompson:2006is,Lacki:2013ry}. In particular, mG
magnetic field strengths have been predicted~\cite{Torres:2004ui} and
measured~\cite{McBride:2015} in the starburst core of Arp 220.  The
cosmic ray population in the starburst is dominated by the nearest
accelerators in time/space to the position of interest, thus breaking
a direct relation between average fields and mean cosmic ray
population~\cite{Torres:2012xk}.  Up to mG field strengths are consistent with the
gamma-ray and radio spectra in the gas-rich starburst cores of NGC 253
and M82~\cite{Paglione:2012ma}.  Besides, the field strength in the
halo of M82 and NGC 253 could be as high as
$300~\mu$G~\cite{DomingoSantamaria:2005qk,delPozo:2009mh,Lacki:2013nda}. Herein
we will remain agnostic with regard to the process responsible for
magnetic field amplification, and we consider all the nearby AGN and
SBG sources which are consistent with Auger and TA observations.

\section{Nucleus Photodisintegration}
\label{sec2}

 The mean free path (mfp) for the different elements is obtained from
 the photodisintegration cross section and the background photon flux
 (of type $k$) as
\begin{equation}
\frac{1}{\lambda_k} =
\frac{1}{2\gamma^2} \, \int_{\varepsilon_{\rm th}/2\gamma}^{\infty} \frac{1}{ \varepsilon^2} \
f_k(\e)
\, d\varepsilon \, \int_{\e_{\rm th}}^{2\gamma \varepsilon} \varepsilon' \,
\sigma_A(\varepsilon') \, d\varepsilon' \,,
\label{rate}
\end{equation}
where $\e_{\rm th}$ is the threshold energy for the reaction in the
nucleus rest frame, $\gamma$ is the relativistic factor for the
nucleus, and $f_k$ is the photon distribution function (number of
photons per unit volume and energy) in the  
frame where the cosmic microwave background (CMB) is at 2.7~K, in
which it is assumed to be isotropic~\cite{Stecker:1969fw}. With a
change of variables $\e\to\e/2\gamma$ we can rewrite (\ref{rate}) as
\begin{equation}
\frac{1}{\lambda_k}=\frac{1}{\gamma}\int_{\e_{\rm
    th}}^\infty\frac{1}{\e^2} \ f_k  \ \left(\frac{\e}{2\gamma}\right) \
{\cal I}(\e)\,  d\e ,
\end{equation}
where \begin{equation} {\cal I}(\e)\equiv\int_{\e_{\rm th}}^\e
  \e' \ \sigma_A(\e') \ d\e'.\end{equation}

For nitrogen, silicon and iron, the cross section is taken from TALYS
1.8 as done by CRPropa3~\cite{Batista:2016yrx}, where the parameters
of the giant dipole resonance (GDR) are modified according to the IAEA
atlas, which show a better agreement with experimental data.  For
helium, the cross section is taken from Eq.~(3) in~\cite{Soriano:2018lly}. The
relevant photon backgrounds are the extragalactic background light
(EBL) and the CMB. For the CMB, we take \begin{equation}
  f_{\rm{CMB}}=\frac{1}{(\hbar
    c)^3}\left(\frac{\e}{\pi}\right)^2\left[e^{\e/T}-1\right]^{-1} \,,\end{equation}
corresponding to a Bose-Einstein distribution with temperature $T =
2.7255(6)~{\rm K}$~\cite{Fixsen:2009ug}.  For the EBL, we take the
results from~\cite{Gilmore:2011ks}.

\begin{figure}[tpb]
\postscript{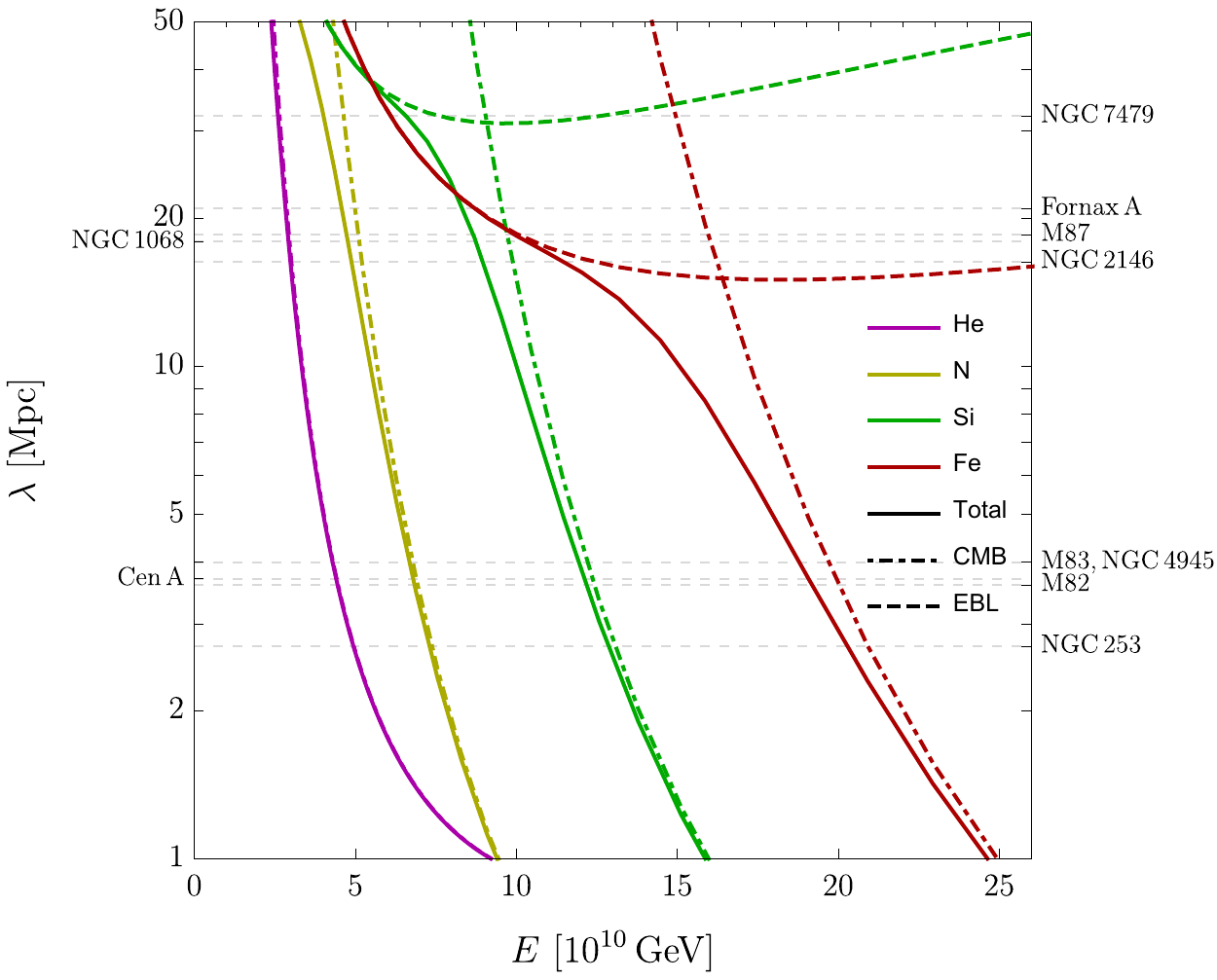}{0.99}
\begin{center}
\caption{Photodisintegration mfp on the CMB and EBL. The horizontal dashed lines indicate the distance to nearby
  starbursts and radio galaxies.
\label{fig:1}}
\end{center}
\end{figure}

In Fig.~\ref{fig:1} we show the photodisintegration mfp for various
nuclei. It is evident that the  mfp
decreases rapidly with increasing energy, and increases rapidly with
increasing nuclear composition. More precisely,
\begin{itemize}
\item at $E = 10^{10.7}$~GeV, the mfp for ionized helium ($^4$He) is
  about 3~Mpc, while at $10^{10.9}~{\rm GeV}$ it is nil;
\item at  $E = 10^{10.9}~{\rm GeV}$, the mfp for ionized nitrogen ($^{14}$N) is about 4~Mpc, while at $10^{11}$~GeV it is nil;
\item at $E =10^{11.1}$~GeV, the mfp for ionized silicon ($^{28}$Si)
  is about 2.5~Mpc, while at $10^{11.2}~{\rm GeV}$ it is nil;
\item  until finally we reach ionized iron ($^{56}$Fe) where
  the mfp at $E = 10^{11.3}$~GeV is about 4~Mpc,
while at $10^{11.4}~{\rm GeV}$ it too is nil.
\end{itemize}
This implies that from sources at increasing distance, fewer and
heavier nuclei at highest energies are expected to reach Earth.  The
main features in the energy evolution of the abundance of various
nuclear species on Earth can be summarized as follows:
\begin{itemize}
\item the contribution of $^4$He should decrease with rising energy and then essentially disappear above
about $10^{10.8}~{\rm GeV}$;
\item on average, only species heavier than $^{14}$N can contribute to the observed flux on
Earth above $10^{11}~{\rm GeV}$, with nuclear species lighter than $^{28}$Si
highly suppressed at $10^{11.2}~{\rm GeV}$;
\item the mean flux of iron nuclei becomes suppressed somewhat below $10^{11.4}~{\rm GeV}$.
\end{itemize}
When the three considerations enumerated above are combined with the
magnetic deflections predicted by (\ref{deflection}) we arrive at the
{\it cepa stratis} structure:
\begin{itemize}
\item in the energy range $10^{10} \alt E/{\rm GeV} \alt 10^{11}$
  light (e.g., $^{4}$He, $^{12}$C, $^{14}$N, $^{16}$O) nuclei are
  expected to survive the trip from nearby (distance $\alt 50~{\rm
    Mpc}$) sources, and these nuclei would suffer average deflections on
  the Galactic magnetic field of $\theta \alt 15^\circ$;
\item at the high energy ($E \agt 10^{11}~{\rm GeV})$ end of the
  spectrum contributions come dominantly from heavier nuclei (e.g.,
  $^{28}$Si, $^{56}$Fe), leading to larger deflection angles
  associated to a decrease in rigidity.
\end{itemize}

To get a rough estimate of the maximum energy observed on Earth we
translate the mfp shown in Fig.~\ref{fig:1} into a cutoff energy in
the spectrum for the various species as a function of the source
distance $D$. The results are listed in Table~\ref{tabla1}.

\begin{table}
\caption{Energy cutoff $E_A (D)$ for various nuclear species. \label{tabla1}}
\begin{tabular}{ccccc}
\hline
\hline
 ~~~$D/{\rm Mpc}$~~~ & ~~~$E_4/{\rm EeV}$~~~ & ~~~$E_{14}/{\rm EeV}$~~~ & ~~~$E_{28}/{\rm
  EeV}$~~~ & ~~~$E_{56}/{\rm EeV}$~~~ \\
\hline
2 to 3 & 60 & 100 & 180 & 220 \\
3 to 4 & 50 & 80 & 130 & 210 \\
16 to 21 & 40 & 60 & 100 & 110 \\
\hline
\hline
\end{tabular}
\end{table}

\section{Deflections on Magnetic Fields}
\label{sec3}

Our understanding of the extragalactic magnetic field strength is
surprisingly vague. Measurements of diffuse radio emission from the
bridge area between the Coma and Abell superclusters~\cite{Kim}
provide an estimate of ${\cal O}(0.2-0.6)\,\mu$G for the magnetic
field in this region, assuming the contributions of the magnetic field
and the relativistic particles are approximately equal (equipartition
condition). Fields of ${\cal O}(\mu{\rm G})$ have also been estimated
in a more extensive study of 16 low redshift
clusters~\cite{Clarke:2000bz}. It is usually conjectured that the
observed $B$-fields result from the amplification of much weaker seed
fields. However, a concrete unified model to explain the initial
weak seed fields is yet to see the light of day. Generally speaking, the models for
the seed fields can be divided into two broad classes: {\it (i)}~{\it
  cosmological models}, in which the seed fields are produced in the
early universe; {\it(ii)}~{\it astrophysical models}, in which the
seed fields are generated by motions of the plasma in
(proto)galaxies. The galactic-scale superwinds generated by the
starbursts provide a particular example of astrophysical models.
Actually, if most galaxies lived through an active phase in their
history, one expects the magnetized outflows from their jets to 
also efficiently pollute the extragalactic medium. It is reasonable to
suspect that the $B$-fields originating in this way would be randomly
oriented within cells of sizes below the mean separation between
galaxies, $\lambda_B \lesssim 1~{\rm Mpc}$.

Thus far the extremely weak unamplified extragalactic magnetic fields
have escaped detection. Measurements of the Faraday rotation in the
linearly polarized radio emission from distant
quasars~\cite{Kronberg:1993vk,Farrar:1999bw} and/or distortions of the
spectrum and polarization properties in the
CMB~\cite{Barrow:1997mj,Jedamzik:1999bm} yield upper limits on the
extragalactic magnetic field strength as a function of the reversal
scale.  It is noteworthy that Faraday rotation measurements (RM)
sample extragalactic magnetic fields of any origin (out to quasar
distances), while the CMB analyses set limits {\em only} on primordial
magnetic fields. The RM bounds are strongly dependent on assumptions
about the electron density profile as a function of the redshift.  If
electron densities follow that of the Lyman-$\alpha$
forest~\cite{Blasi:1999hu}, the average magnitude of the magnetic
field receives an upper limit of $B \sim 0.65$~nG for reversals on the
scale of the horizon, and $B \sim 1.7$~nG for reversal scales on the
order of 1~Mpc, at the $2\sigma$ level~\cite{Pshirkov:2015tua}. These
upper limits are estimated assuming standard cosmological
parameters~\cite{Tanabashi:2018oca}.

In the limit of small deflections (expected for nG field strength) the
typical deflection of UHECRs in the extragalactic magnetic field can
be estimated to be
\begin{equation}
\theta \approx 0.15^\circ Z \  \sqrt{\frac{D}{3.8~{\rm Mpc}} \ \frac{\lambda_B}{0.1~{\rm Mpc}}} \ \left(\frac{B}{{\rm nG}}
\right) \ \left(\frac{10^{11}~{\rm GeV}}{E} \right) \,,
\end{equation}
where $D$ is the source distance and $Z$ is
the charge of the UHECR in units of the proton
charge~\cite{Waxman:1996zn,Farrar:2012gm}. It is then reasonably to
assume that extragalactic deflections would generally be much smaller
than those arising from the galactic magnetic field (GMF).

We now turn to study the effect of GMF on the deflection of UHECRs
nuclei from the direction of nearby starburst and radio galaxies. We
take the Jansson and Farrar (JF) model as a semi-realistic magnetic
field model to investigate deflections as UHECRs travel through the
Galaxy~\cite{Jansson:2012pc,Jansson:2012rt,Unger:2017kfh}. In the JF
model the GMF is described by a superposition of three divergence-free
large-scale regular components: a spiral disk field, a toroidal halo
field, and a poloidal field. In addition, there is a turbulent random
magnetic field that follows a Kolmogorov distribution.

All the magnetic fields are implemented in the Runge-Kutta cosmic ray
propagation simulation tool {\em CRT}~\cite{Sutherland:2010vx}.  The
random magnetic field is produced within a cubic box of side
$5.12\,{\rm kpc}$. Inside this box, a different value of the magnetic
field is produced in each of $512^3$ equally spaced points. This box
is then replicated and placed through the galactic space with
different orientations, to cover the whole galaxy. For the random
magnetic field, which is described as a superposition of waves of
different wavelengths, we restrict them to range from $5\,{\rm pc}$ to
$30\,{\rm pc}$.

The parameters of the regular and random magnetic field are
constrained by: {\it (i)}~multi-frequency radio observations of the
Faraday RM of extragalactic radio sources; {\it (ii)}~measurements of
the polarized synchrotron emission of cosmic ray electrons in the regular
magnetic field of the Galaxy; {\it (iii)}~measurements of the total
(polarized and unpolarized) synchrotron intensity, which is a
line-of-sight integral depending on the product of cosmic ray electron density
and total transverse magnetic field strength (coherent and random).

\begin{figure*}[tbp]
\begin{minipage}[t]{0.95\textwidth}
    \postscript{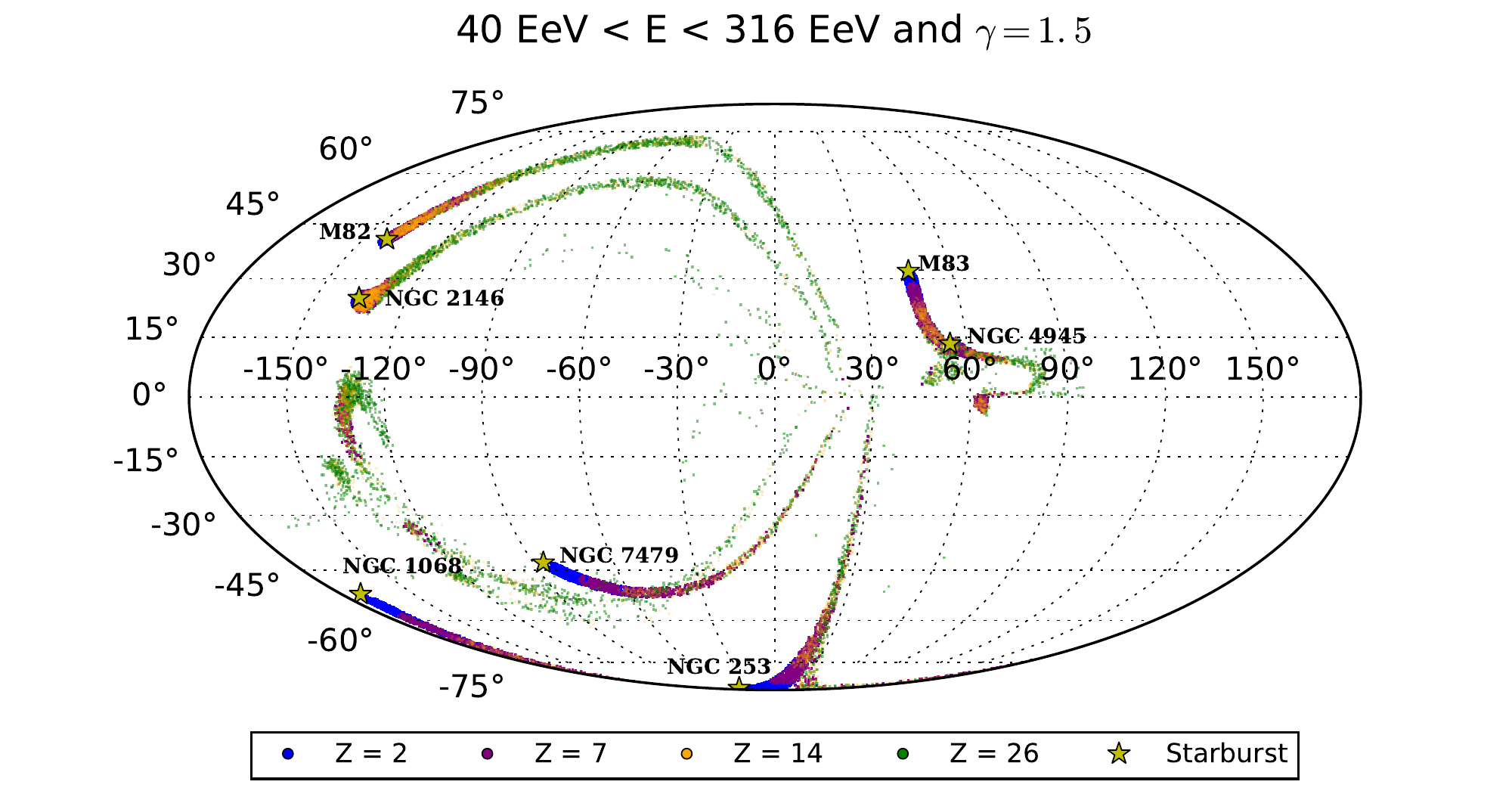}{0.85}
\end{minipage}
\hfill \begin{minipage}[t]{0.95\textwidth}
  \postscript{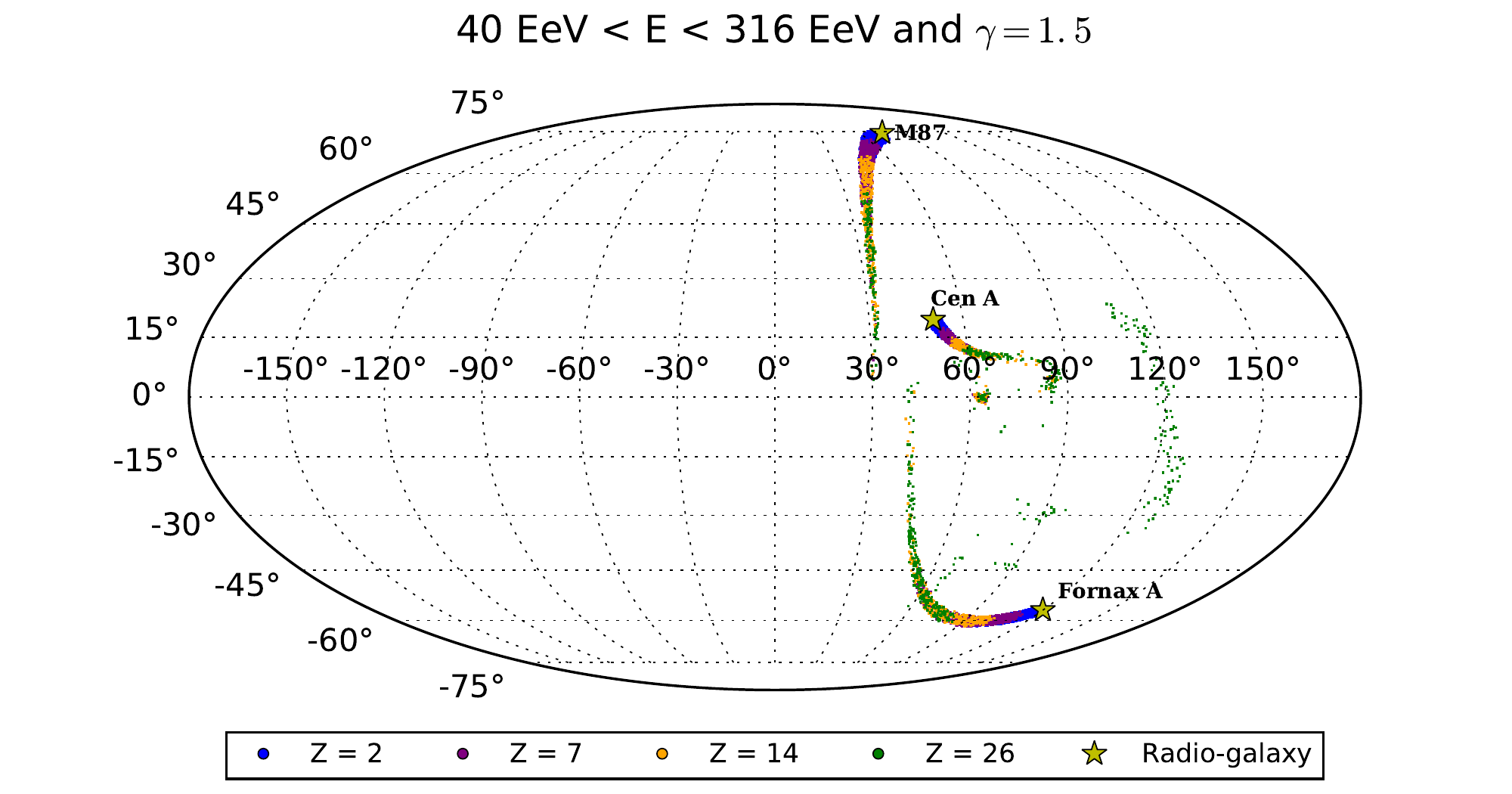}{0.85}
\end{minipage}
\caption{Skymaps in Mollweide projection of the distribution of
  arrival directions for selected SBGs (up) and radiogalaxies (down). The sky maps are in
    Galactic coordinates.}
\label{fig:nocuts}
\end{figure*}

\begin{figure*}[tbp]
\begin{minipage}[t]{0.95\textwidth}
    \postscript{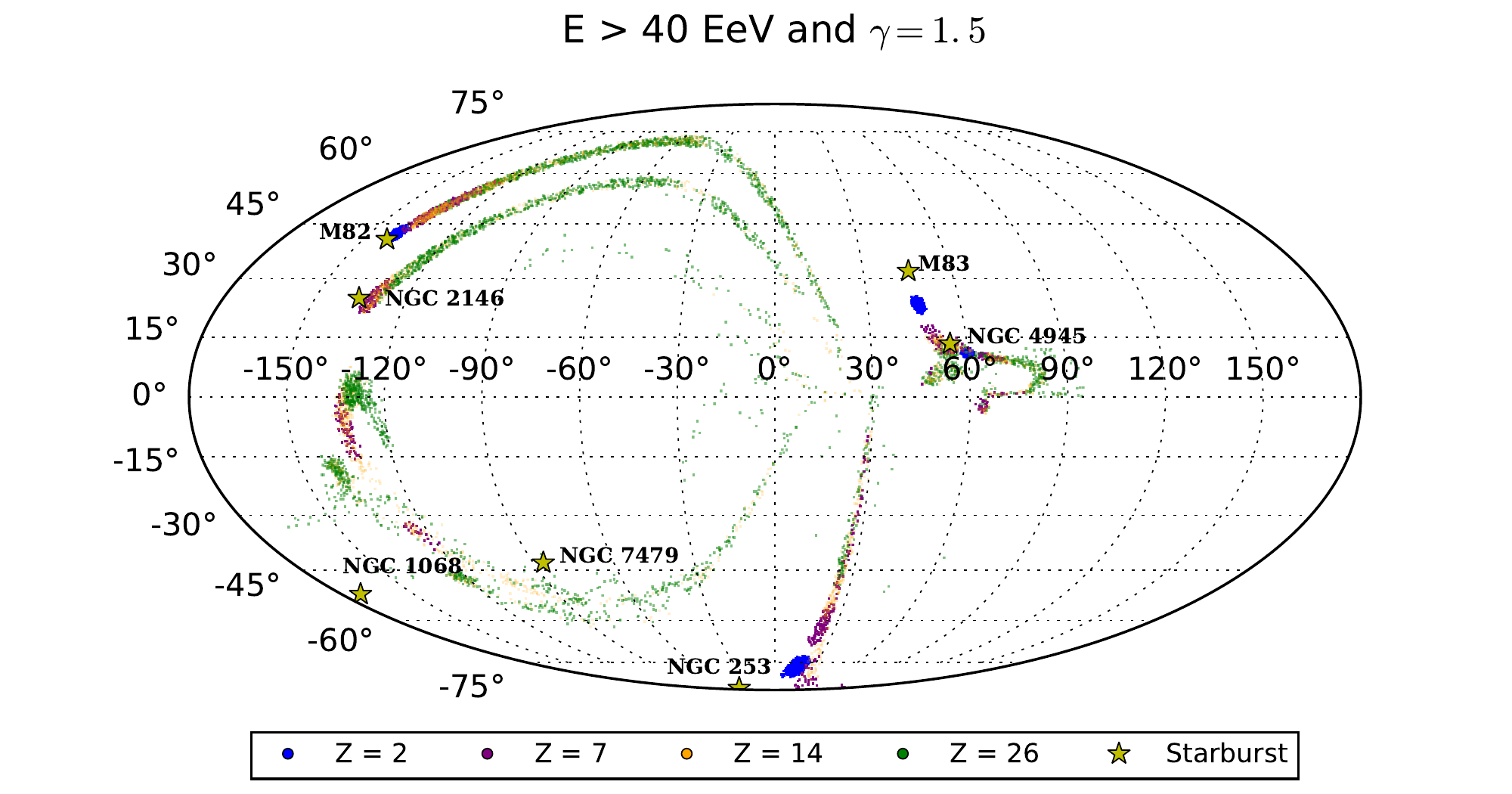}{0.85}
\end{minipage}
\hfill \begin{minipage}[t]{0.95\textwidth}
  \postscript{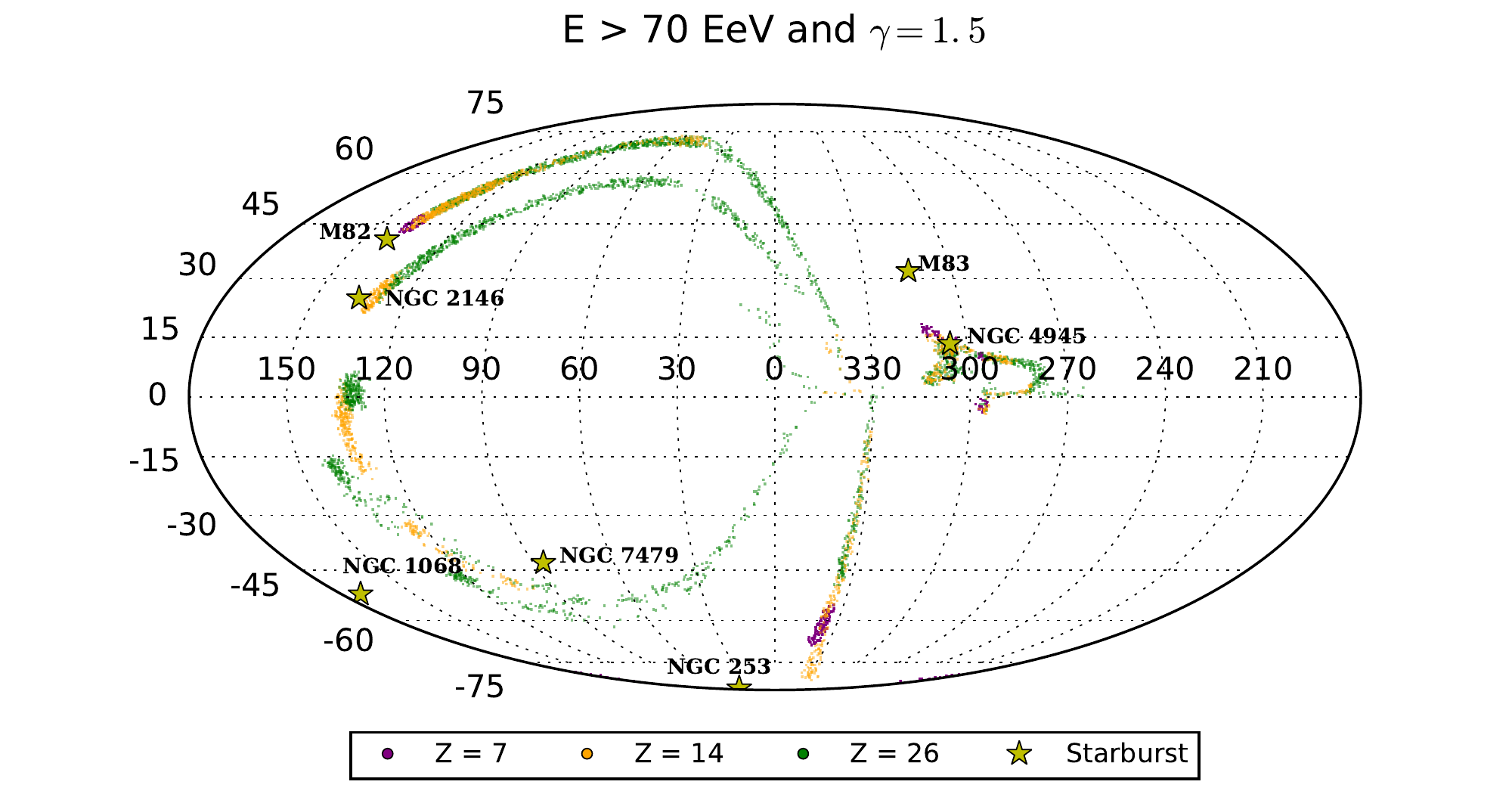}{0.85}
\end{minipage}
\begin{minipage}[t]{0.95\textwidth}
    \postscript{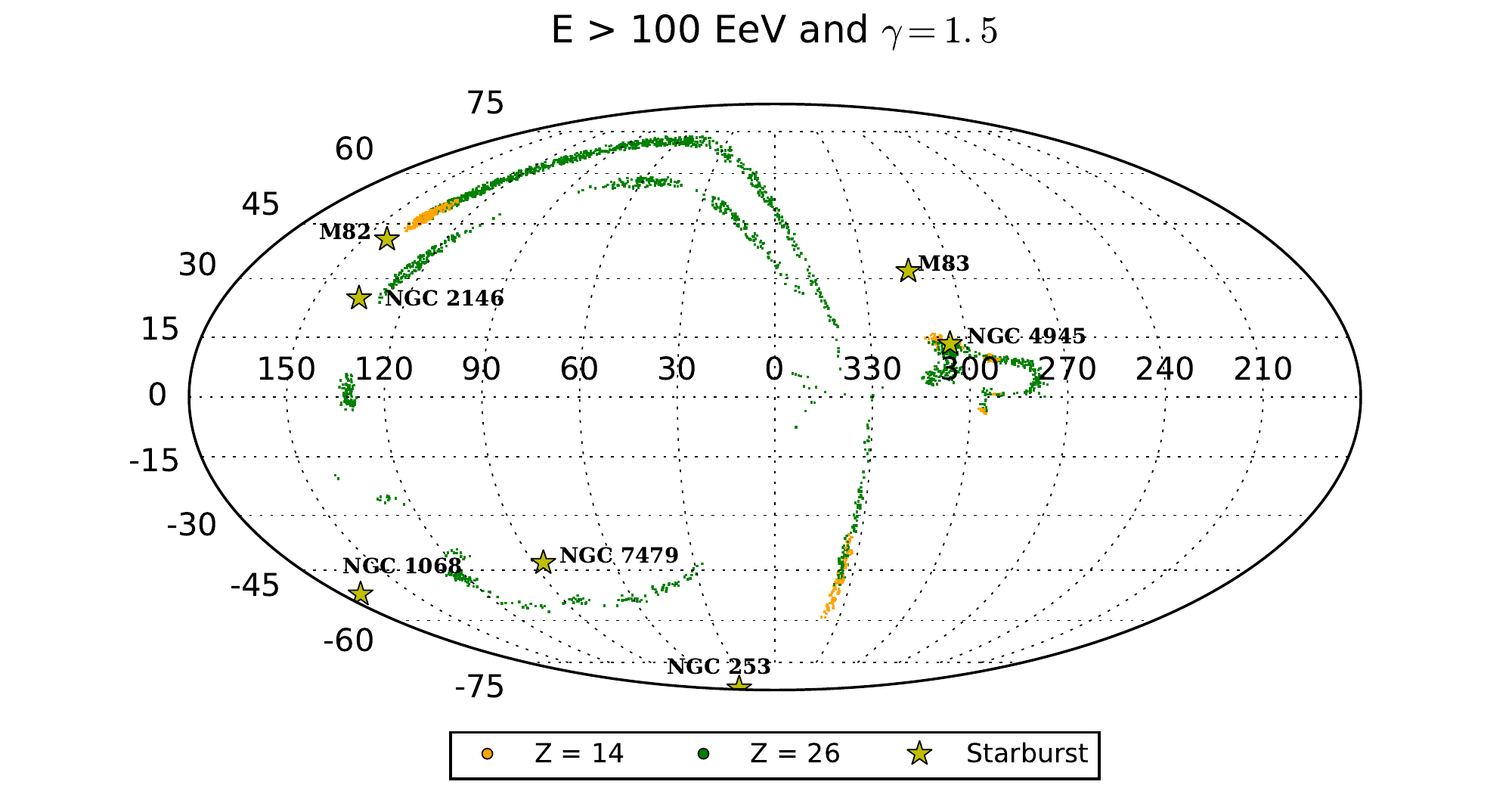}{0.85}
\end{minipage}
\caption{Skymaps in Mollweide projection of the distribution of
  arrival directions for selected starbursts shown in Fig.~\ref{fig:1}
  here indicated with yellow star. In all the cases we adopted a hard injection spectrum
  $\propto E^{-1.5}$, setting a threshold for the minimum energy of
  $E_{\rm min}/{\rm EeV}
    = 40,\, 70,\, 100$ from top to bottom. We have also imposed the
    cuts given in Table~\ref{tabla1}. The sky maps are in
    Galactic coordinates.}
\label{fig:skymap1}
\end{figure*}
\begin{figure*}
\begin{minipage}[t]{0.95\textwidth}
  \postscript{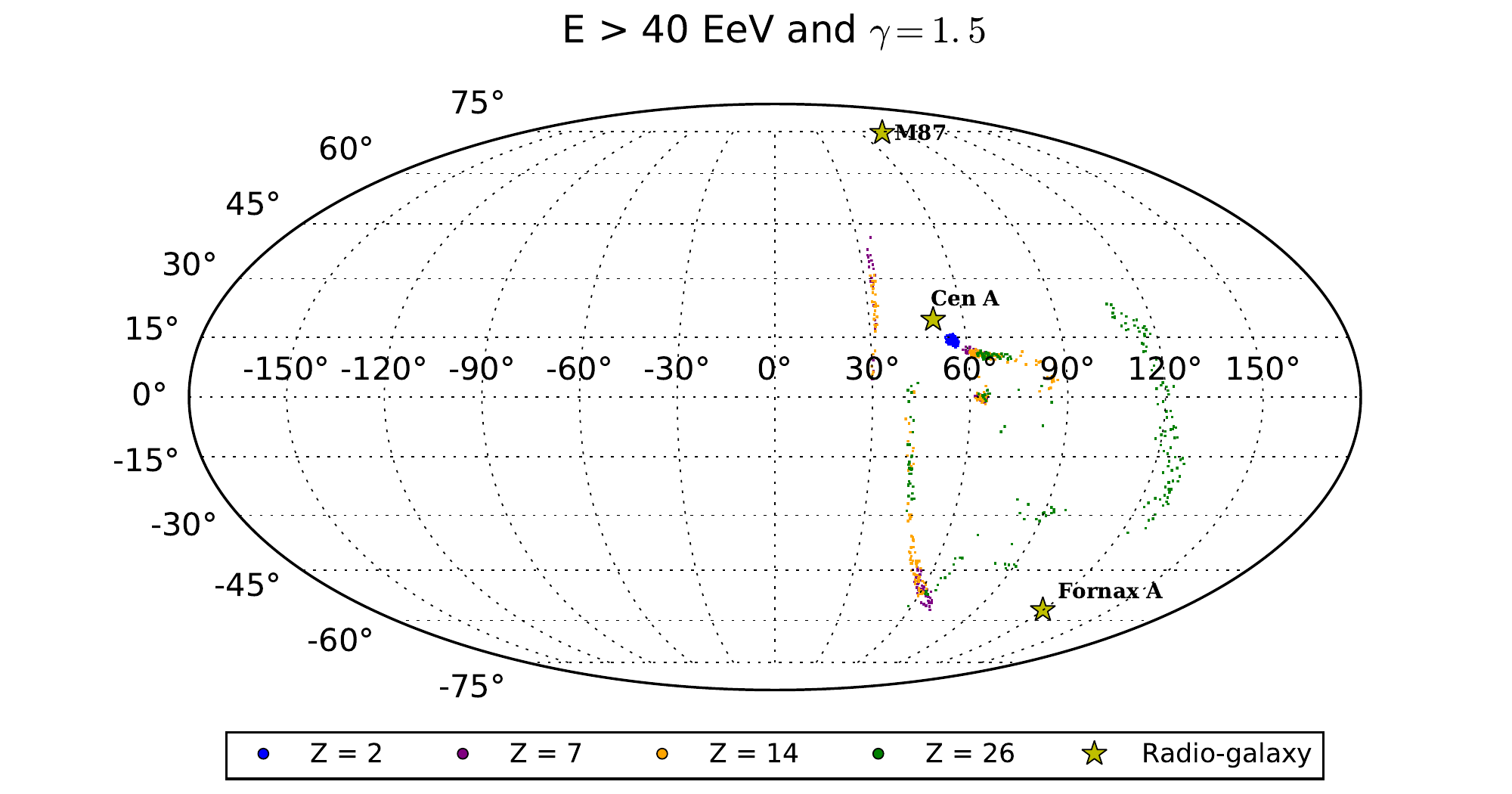}{0.85}
\end{minipage}
\begin{minipage}[t]{0.95\textwidth}
    \postscript{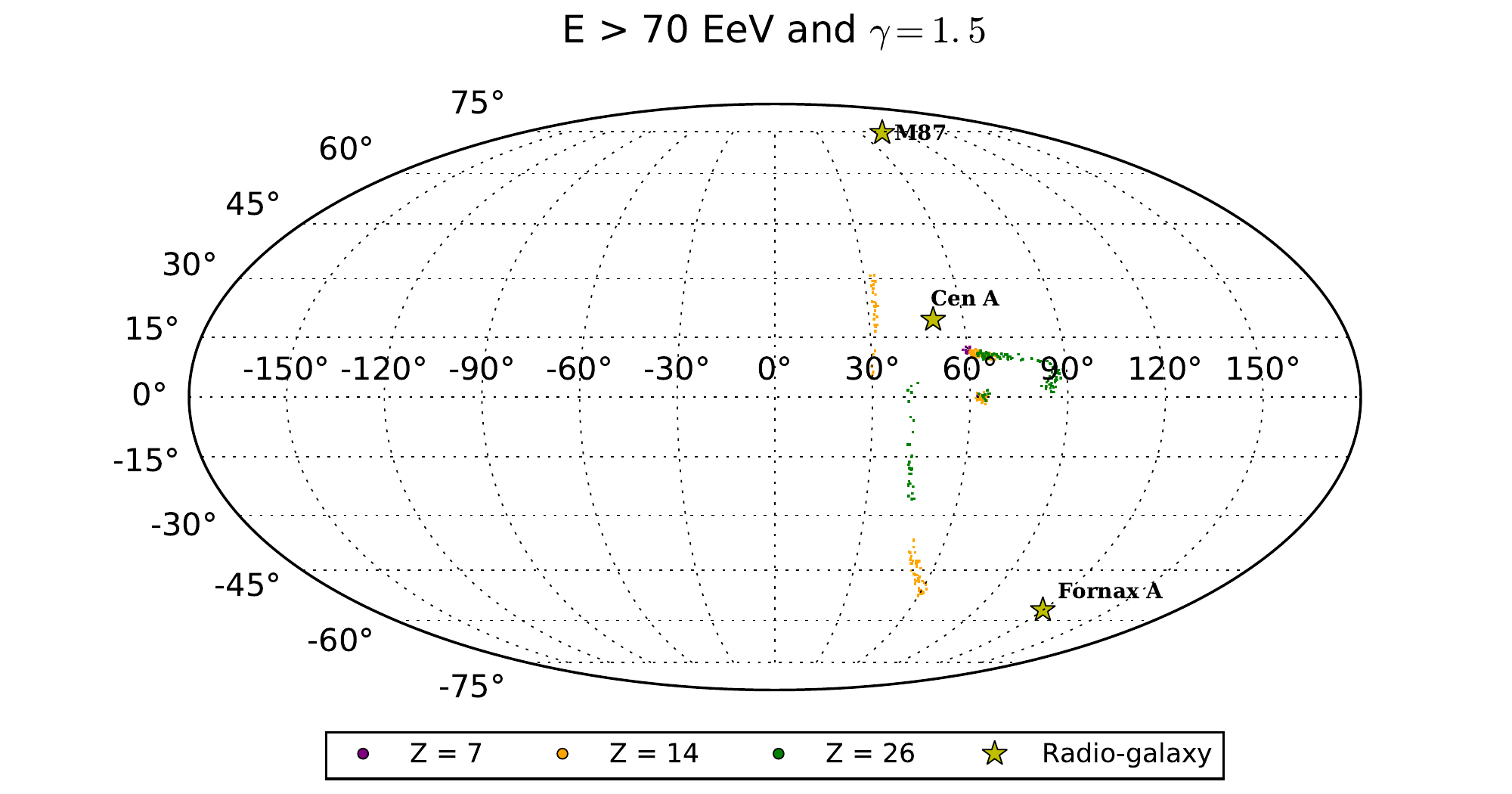}{0.85}
\end{minipage}
\hfill \begin{minipage}[t]{0.95\textwidth}
  \postscript{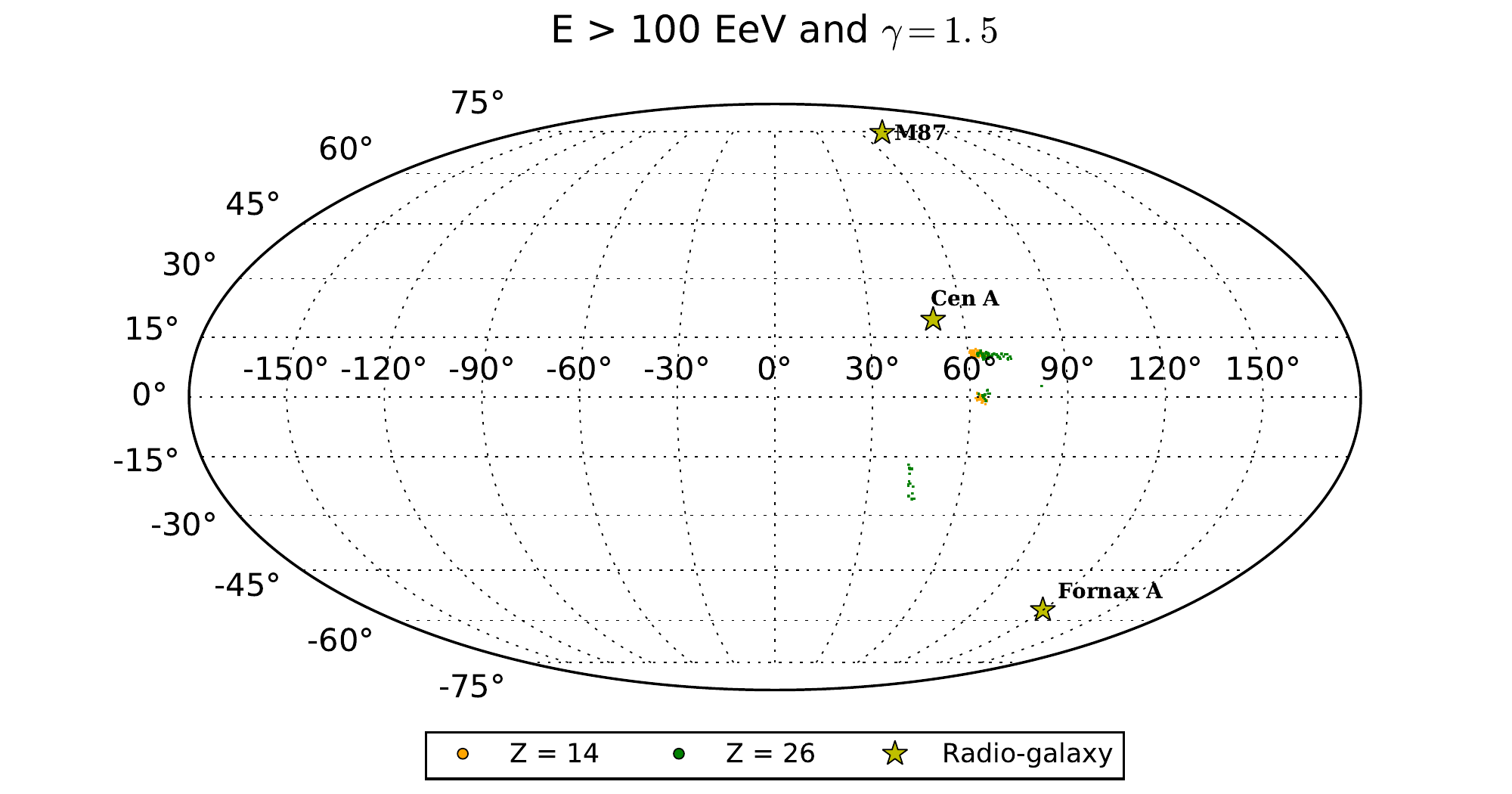}{0.85}
\end{minipage}
\caption{Skymaps in Mollweide projection of the distribution of
  arrival directions for selected 
  radio galaxies also shown in Fig.~\ref{fig:1} here indicated with a
  yellow star. In all the cases we adopted a hard injection spectrum
  $\propto E^{-1.5}$, setting a threshold for the minimum energy of
  $E_{\rm min}/{\rm EeV}
    = 40,\, 70,\, 100$ from top to bottom. We have also imposed the
    cuts given in Table~\ref{tabla1}. The sky maps are in Galactic coordinates.}
\label{fig:skymap2}
\end{figure*}

Beginning with an isotropic distribution of arrival directions
observed on Earth we back-propagate $10^6$ nuclei to the border of the
Galaxy for each of the species {$^4$He}, {$^{14}$N}, {$^{28}$Si}, and
{$^{56}$Fe}.  Although we have shown in Table~\ref{tabla_gamma} that in
the energy range of interest the source spectra are rather soft, to
illustrate the competition between energy loss during propagation and
deflection on the GMF in a simple way we adopt hard energy spectra
$\propto E^{-1.5}$, with a maximum energy $E_{\rm max} =
10^{11.5}~{\rm GeV}$.  We tally the fraction of events consistent with
the directions in the sky of nearby starburst and radio galaxies.  A
summary of the corresponding deflections exhibited as Mollweide
projections is shown in Figs.~\ref{fig:nocuts}, \ref{fig:skymap1} and
\ref{fig:skymap2}.  In Figs.~\ref{fig:skymap1} and \ref{fig:skymap2}
we have imposed the energy cuts for the different species given in
Table~\ref{tabla1} to account for the energy loss before reaching the
Galaxy. One can draw the following conclusions:
\begin{itemize}
\item Our results are consistent with similar analyses using the same
  JF model and the hypotheses that M87 and Fornax
  A~\cite{Smida:2015kga}, or Cen A and M82~\cite{Farrar:2017lhm} are
  potential sources of UHECRs.
\item We observe that
  the effect of the GMF is to modify the onion-like structure one
  would expect if there were purely random magnetic fields into more
  complex elongated banana shapes.
\item A comparison of Fig.~\ref{fig:nocuts}  with the observed excess map of Fig.~\ref{fig:0.5}  indicates qualitative agreement with Auger data.
\item It was proposed that Fornax A could explain the bulk of the
  Auger warm spot right of the Galactic south
  pole~\cite{Matthews:2018laz}. However, as shown in Fig.~\ref{fig:skymap2}, the JF
  model predicts deflections which do not favor this association.  
\item A comparison of Figs.~\ref{fig:0.5} and \ref{fig:skymap1} shows
  that the JF model with selected turbulent parameters cannot explain
  the TA hot spot. However, the GMF in this region is dominated by
  turbulence (see e.g., Fig.~11 in \cite{Farrar:2017lhm}), which can
  accommodate an abundance of possible sky patterns that could be
  consistent with TA observations.
\end{itemize}
In the remainder of the paper we develop a test which a future mission
such as POEMMA could use to clarify the nuclear composition of a given
hot spot.

\section{Test Statistics}

\label{sec4}

We have seen that UHECRs coming from a given source in the sky are
scattered around the line of sight to that source. Their arrival
directions depend on the properties of the cosmic ray, as well as on the
intergalactic and galactic media. Although the GMF is highly
anisotropic, we will assume that the deflection of particles is
isotropic around the line of sight. In general, one must consider the
variations of the magnetic field for UHECRs arriving from different
points of the sky. The anisotropies are rather large, as shown in
Sec.~\ref{sec3}. A full consideration of the anisotropic magnetic
deflections would modify the distribution (\ref{pdf}) below to include
an azimuthal variable around the line of sight, and should also take
into account its direction in the sky. Nevertheless, the procedure of
the analysis would not change significantly. The assumption of
isotropy around the line of sight allows us to demonstrate the search
technique while keeping the complexity at a reasonable level at this
stage.

Hereafter we assume that the magnitude of the deflection of a cosmic
ray, with energy $E$ and charge $Ze$, about the line of sight is given
by (\ref{deflection}). With this simplified picture of the effect that
magnetic fields have on UHECRs, one can assume that cosmic rays are
\emph{normally} distributed around the source direction, which defines
the center of the hot spot. The generalization of a normal
distribution to directional data is the wrapped normal distribution,
which can be approximated by the von Mises
distribution~\cite{Anchordoqui:2018qom}. In what follows we consider
that the deflection $\delta$, which characterizes the angle between
the arrival direction and the line of sight, to be a random variable
distributed according to a one sided von Mises distribution,
bounded by a window size $\Delta$ with zero mean and a dispersion
parameter $\kappa=1/\theta^2(E,Z)$. Thus, its probability density
function is \begin{equation} f_{\mathrm{vM}}(\delta|E,Z)\propto
  \exp\left[\frac{\cos\delta}{\theta^2(E,Z)}\right] \
  \Theta(\Delta-\delta).\end{equation}

The observed UHECR spectrum can be described as being proportional to
$\sum_A \sum_s w_{A,s} \ E^{-\gamma}\exp[-E/E_{A,s}(D)]$, where
$E_{A,s}(D)$ is the cutoff energy that depends on the baryon number of
the nucleus and the distance to its source; see Table~\ref{tabla1}. Here, the weights
$w_{A,s}$ account for the various contributions of different species
$A$ for a given source $s$.  Moreover, a lower cut in the energies of
interest is considered. Therefore, the probability density for a
cosmic ray assumed to come from a distant point source to have energy
in $[E,E+ dE]$ and deflection in $[\delta,\delta+
d\delta]$ is
\begin{widetext}
\begin{equation}
f(E,\delta|A,Z,z,\Delta,E_0) =\mathcal A
 \ E^{-\gamma} \ \exp\left[-\frac{E}{E_A(z)}\right] \
 \exp\left[\frac{\cos\delta}{\theta^2(E,Z)}\right] \ \Theta(\Delta-\delta)
 \ \Theta(E-E_0),\label{pdf} 
\end{equation}
\end{widetext}
where $\mathcal A$ is a normalization constant. Since this
distribution represents the measured spectrum, and not the actual
spectrum at Earth, it must be understood that this probability density
represents the distribution of the events the experiment would
record. If one wants to model the actual spectrum following
(\ref{pdf}), it would be necessary to add an energy dependent function
modeling the response of the detector at different energies. In a
full study, this could be easily implemented without changing the
techniques described below.

\begin{figure}[h]
\postscript{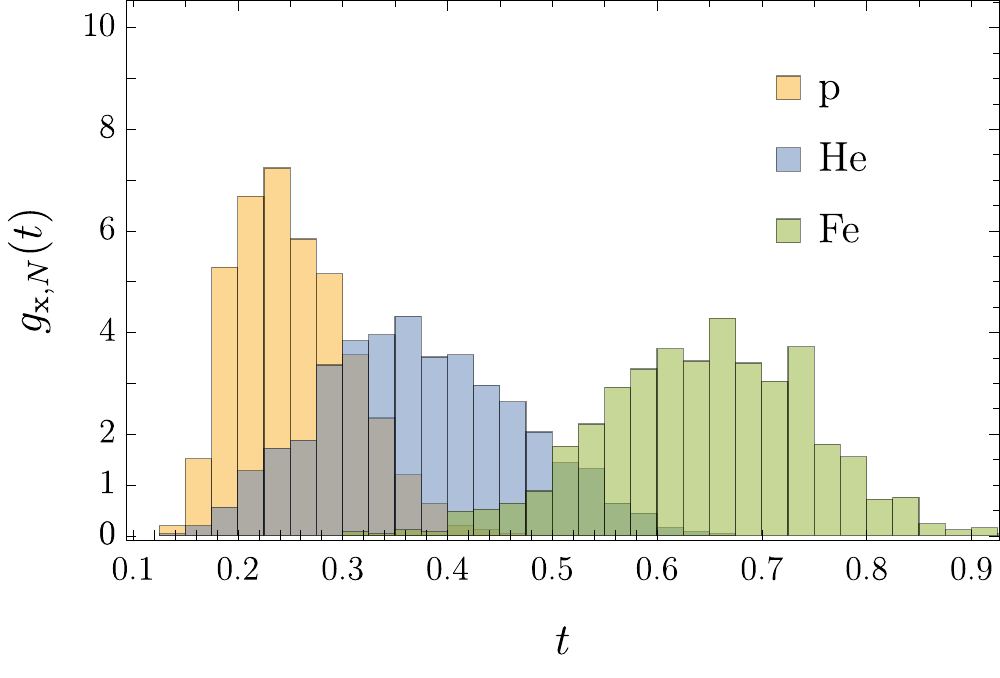}{0.99}
\caption{PDF histograms for the test statistic empirical distribution
  for the null hypothesis (proton), helium and iron, with $N=20$. \label{fig:histograms} }
\end{figure}

Once experimental data is collected, and a series of events within an
angular window are selected to belong to some cosmic ray source, one can study
their energies and deflections to extract, by means of (\ref{pdf}),
information about the composition of the source. Given the theoretical
distributions presented above for different atomic numbers,
statistical testing of the data will provide this information.

To carry out the statistical analysis we must first define the window
size $\Delta$ and the threshold energy $E_0$. Next, the source must be
identified, \emph{i.e.} we select $D$. After that we can calculate the
likelihood that different probability distributions (for different
nuclei) describe the data. This would allow an estimation of the
nucleus producing the major contribution to a given hot
spot. Furthermore, one could use the likelihood ratio for different
nuclei as a tool to study our ability to distinguish them. In general
terms, it is possible to propose a null hypothesis (\emph{e.g.}, that
the composition is only protons), simulate data following the null
hypothesis (${\cal H}_0$) and, choosing a convenient test statistic,
study its distribution for the generated data. Once real data are 
available, the value of the test statistic for that data will provide
a way to test the null hypothesis.

The Kolmogorov-Smirnov (KS) test provides a computationally less
expensive test statistic than the one coming from likelihood
minimization. It allows for the comparison of empirical multivariate
distributions to statistical models, and provides a method for judging
to which extent some data is likely to follow a given statistical
distribution. Given a set $\mathcal D$ of (empirical or simulated)
data points, it is possible to construct an empirical cumulative
distribution function (CDF) $\tilde F(E,\delta)$, which counts the
fraction of data points with energy below $E$ and deflection below
$\delta$.\footnote{More generally, one could consider CDFs defined as
  the fraction of events above $E$ and $\delta$, or combinations of
  \emph{above} and \emph{below} for both variables. We do not study
  those cases here, without denying their relevance.} The CDF for the
null hypothesis is \begin{equation}F(E,\delta|{\cal H}_0)=\int_{E_0}^E
  \ d E' \int_0^\delta \ d\delta'\,f(E',\delta'|{\cal H}_0).\end{equation}
The KS test statistic for $\mathcal D$ is \begin{equation}
  t=\sup_{E,\delta}|F(E,\delta|{\cal H}_0)-\tilde
  F(E,\delta)|,\label{ts}\end{equation} where $E\in[E_0,\infty]$ and
$\delta\in[0,\Delta]$. If each dataset is simulated several times
following the same statistical distribution, one can obtain not only a
single value for $t$, but a distribution $g(t)$ for its value. These
distributions coming from different datasets will give
information on the ability of the experiment and the test to probe a
hypothesis.

The power of a statistical test is the probability that the null
hypothesis is rejected if it is actually false. It is dependent on the
significance level of the test $\alpha$, the probability of rejecting
the null hypothesis while it is true. For a chosen null hypothesis
${\cal H}_0$ and significance level $\alpha$, there is a critical value for
the test statistic, $t_c$, above which there is a fraction $\alpha$ of
the data simulated following ${\cal H}_0$. For a given alternative hypothesis
${\cal H}_k$, the fraction $\beta_k$ of the data with test statistic $t<t_c$
is the probability of not rejecting the null hypothesis while it is
false.  Thus, the power of the test for a given alternative hypothesis
is given by \begin{equation}\mathcal P_k=1-\beta_k=1-\int_0^{t_c}
  g_k(t) \ dt.\label{eq:power}\end{equation}

To exemplify this method, we simulate datasets $\mathcal D_{x,N}$
following the distributions in (\ref{pdf}), where $x\in\{p, {^4{\rm
    He}}, {^{14}{\rm N}}, {^{56}{\rm Fe}}\}$ and $N=\dim\mathcal
D_{x,N}$ is the number of data points in the hot spot. We assume
\emph{(i)} an angular window of $\Delta =13^\circ$, \emph{(ii)} a
distance to the source of about $4\,{\rm Mpc}$, similar to that of
many of the sources considered above, \emph{(iii)} an energy threshold
at $E_0=40\,{\rm EeV}$, \emph{(iv)} and an exponent $\gamma=5.03$,
consistent with both the energy spectrum above $40\,{\rm EeV}$
reported by the Auger Collaboration~\cite{Aab:2017njo} and the source
spectra given in Table~\ref{tabla_gamma}. For each value of $N$, which
roughly corresponds to a given life time of the experiment, we
consider as null hypothesis a pure proton composition, ${\cal
  H}_{\mathrm p,N}$, and the different nuclei as alternative
hypotheses ${\cal H}_{\mathrm x,N}$.

\begin{figure}[h]
\postscript{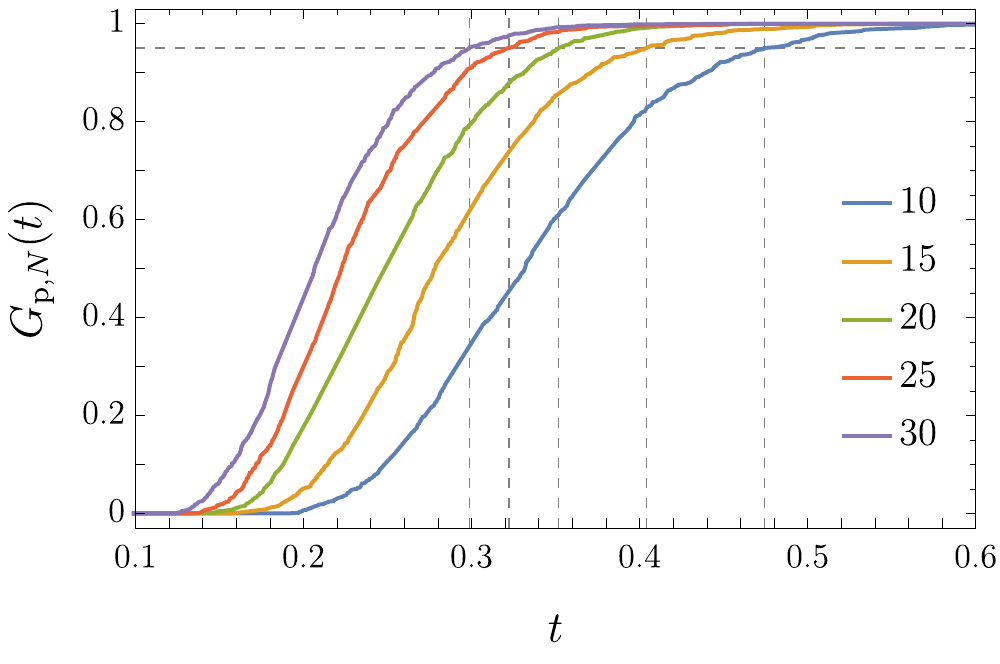}{0.99}\caption{CDF for the test statistic
  distribution for null hypotheses for various $N$. The dashed lines
  indicate the choice $\alpha=0.05$ and the corresponding critical
  values of the test statistic. \label{fig:nullcdf}}
\end{figure}

Each dataset is simulated $10^3$ times to obtain the test statistic
distributions. Some of them are shown in Fig.~\ref{fig:histograms}. The ability to distinguish the null from the alternative hypotheses decreases with the overlap of the different distributions.  In Fig.~\ref{fig:nullcdf} we show the CDF $G_{\mathrm{x},N}(t)=\int_0^tg_{\mathrm x,N}(t')\ dt'$ for protons to illustrate how the choice of $\alpha$ provides the critical values of $t$ as $G_{\mathrm{p},N}(t_{c,N})=1-\alpha$. Introducing the CDF in (\ref{eq:power}), the power is given as $\mathcal P_{\mathrm{x},N}=1-G_{\mathrm{x},N}(t_{c,N})$.

\begin{figure}[h]
 \postscript{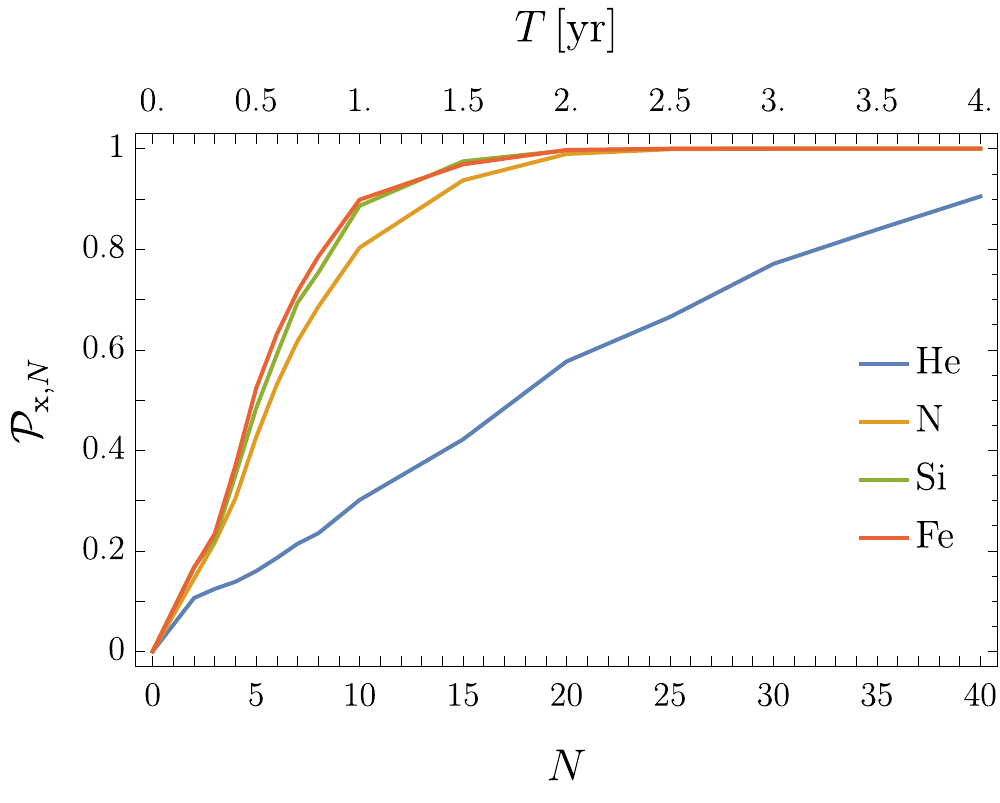}{0.99}\caption{Power of the statistical test
   for different alternative hypotheses, \emph{i.e.} different nuclei
   and number of events per hot spot. The horizontal axis on the top
   indicates the projected time-scale for POEMMA. \label{fig:powers}}
\end{figure}

In Fig.~\ref{fig:powers} we show the statistical power of the test
considering different alternative hypotheses, as a function of
$N$. If the hot spot is composed of nuclei heavier than nitrogen,
observation of $N\agt 20$ events will be required to discard a pure-proton
explanation at the 95\% CL.

\begin{figure}[h]
 \postscript{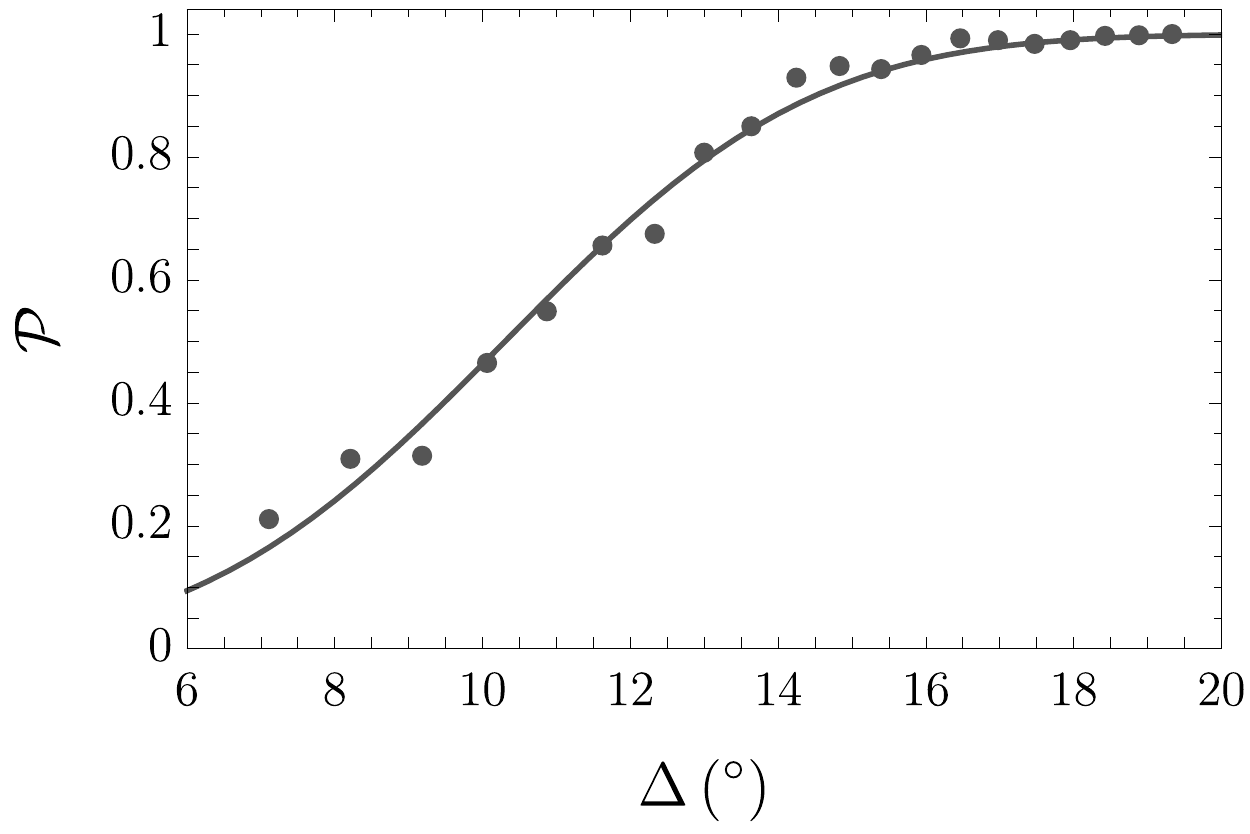}{0.99}\caption{Power of the statistical test
   for a nitrogen alternative hypothesis as a function of the angular radius of the window around the source. The number of events in the sky is fixed to have an expected number of 10 events in a $13^\circ$ radius window. The curve shows a fit to the data with an error function. \label{fig:window}}
\end{figure}

The variation of the statistical power with the radius of the angular window is presented in Fig. \ref{fig:window} for the case of nitrogen, keeping constant the number of events per steradian. 

\begin{figure}[h]
 \postscript{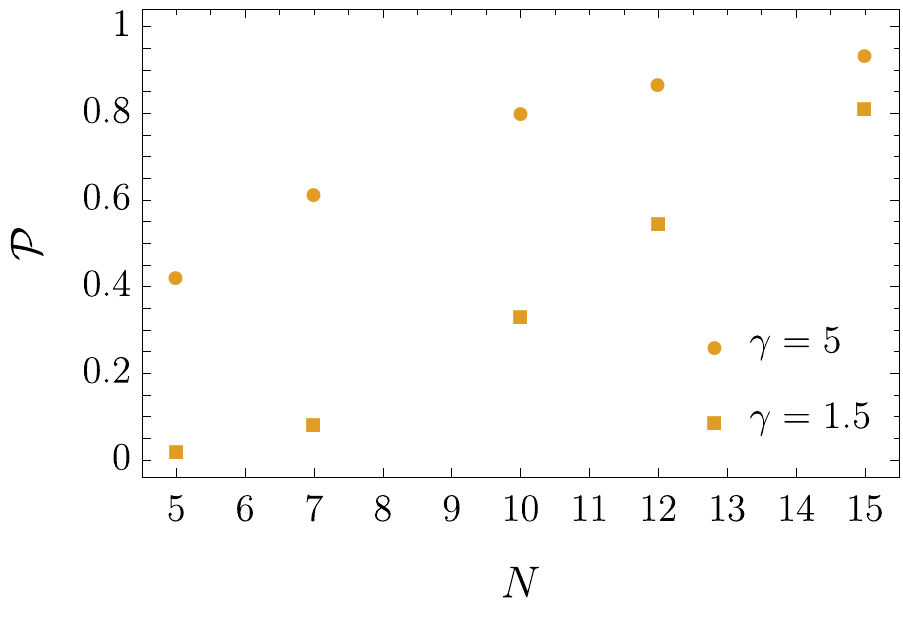}{0.99}\caption{Power of the statistical test
   for a nitrogen alternative hypothesis for a hard ($\gamma=1.5$) and a soft ($\gamma=5$) spectrum, as a function of the number of events in the hotspot. All the other parameters are unaltered with respect to those in Fig. \ref{fig:powers}.}\label{fig:gamma1.5}
\end{figure}

We have briefly study the effects of considering a harder spectrum. This increases the number of high energy nuclei, making the distributions of energy and arrival direction resemble more those of protons. Overall, the test statistic distributions for nuclei shift to lower values, increasing the overlap with the proton distributions. In Fig. \ref{fig:gamma1.5} we show the comparison between the statistical power for soft and hard spectra for nitrogen.

It should be clarified that this method is not meant to obtain the most probable composition of the source, nor to obtain the contributions of different nuclei to a given hotspot. We illustrate a method to reject a pure composition scenario. In order to obtain a more detailed information about the composition of the hotspot, it would be possible to add several terms like (\ref{pdf}) with different weights for different nuclei, and estimate the values of the weights from the data. Nevertheless, it can be safely stated that the power to distinguish a mixed composition sample from a pure proton composition with this method will never be lower than that to distinguish the lightest nuclei in the sample from protons. In any other situation (trying to reject a nuclei pure composition or a mixed composition), the power will decrease with respect to that presented here, as the overlap between the test statistic distributions will increase.

\section{POEMMA Sensitivity}
\label{sec5}

\begin{figure}[h]
 \postscript{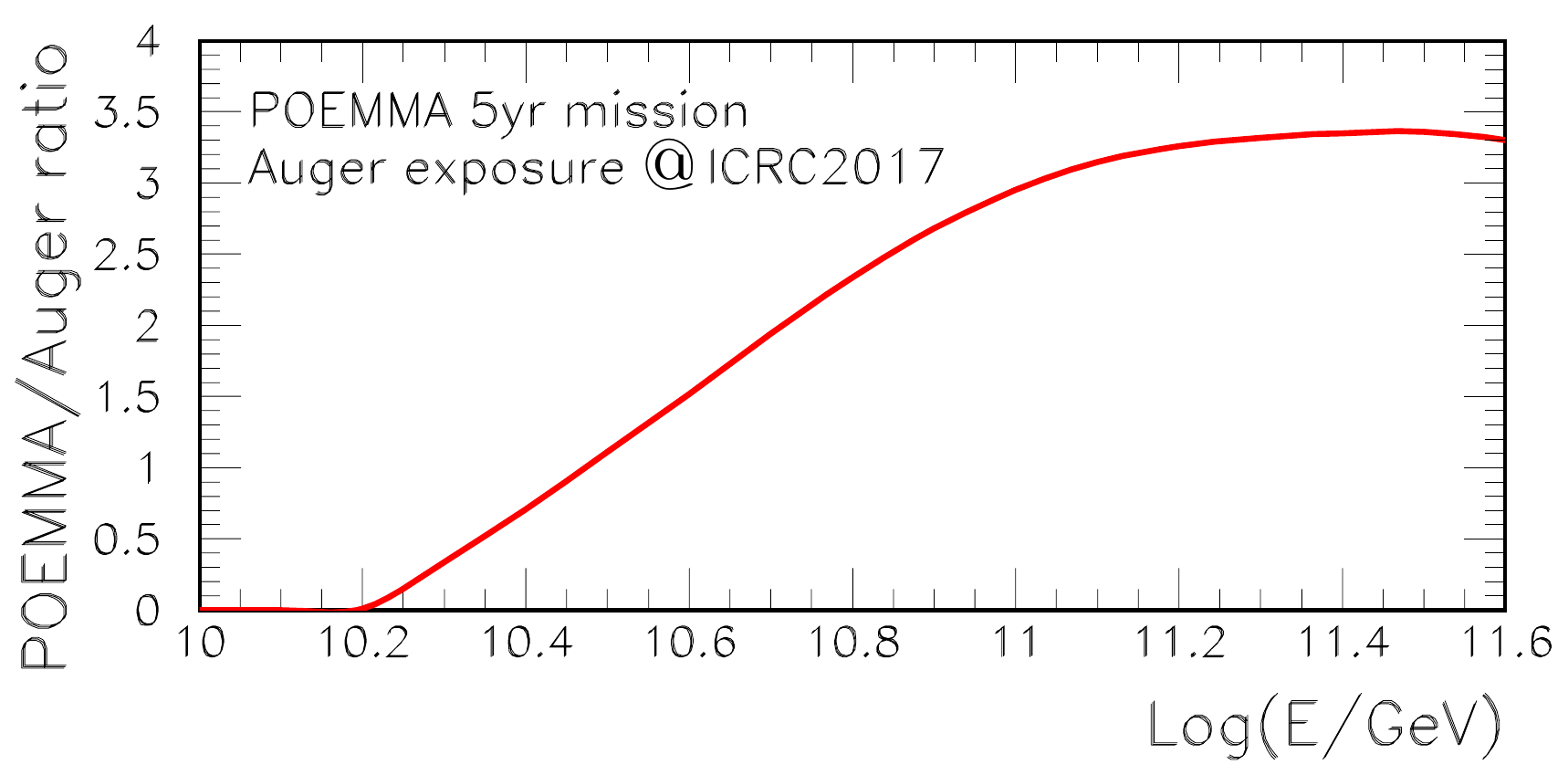}{0.99}\caption{Ratio of the expected
   exposure of POEMMA after 5 yr of operation and the exposure
   collected by the surface array of the Pierre Auger Observatory as
   reported at the 35th International Cosmic Ray Conference (ICRC 2017)~\cite{Aab:2017njo}.\label{fig:poemma}}
\end{figure}

The NASA's POEMMA mission design~\cite{Olinto:2017xbi} combines the
concept developed for the Orbiting Wide-field Light-collectors
(OWL)~\cite{Stecker:2004wt} mission and the recently proposed
CHerenkov from Astrophysical Neutrinos Telescope
(CHANT)~\cite{Neronov:2016zou} concept to form a multi-messenger probe
of the most extreme environments in the Universe. Building on the OWL
concept, POEMMA is composed of two identical satellites flying in
formation with the ability to observe overlapping regions during
moonless nights at angles ranging from nadir to just above the limb of
the Earth. For a rough estimate of the expected event rate, we
consider the orbit of POEMMA at an altitude 525~km with a separation
between satellites of 300~km each with a field of view of
$45^\circ$. The area observed in stereo at nadir is approximately
$1.46 \times 10^5~{\rm km}^2$, yielding an instantaneous aperture
$\sim 4.6 \times 10^5~{\rm km}^2 \, {\rm sr}$. Preliminary studies on
trigger efficiency and the optical performance of POEMMA indicate the
detector will be fully efficient above about $10^{11}~{\rm GeV}$. We
define the acceptance conditions such that the background from airglow
in the entire focal plane produces a rate below 1~kHz. We require 
signal above threshold in both satellites. Herein we estimate the
expected number of events by scaling the number observed at Auger
according to the ratio of the POEMMA to Auger exposures. More
precisely, in Fig.~\ref{fig:poemma} we compare the exposure to be
collected in 5~yr by POEMMA, assuming a conservative 10\% duty cycle,
with the exposure collected by the Auger surface array as reported
in~\cite{Aab:2017njo}.  The ratio of the exposures is roughly an order
of magnitude larger when comparing with the data collected by the
fluorescence detectors of Auger. The ratio of the number of events
(POEMMA vs Auger, bin by bin of energy) is readable from the exposure
scaling.

In order to make estimations about the future performance of POEMMA
for the task presented in Sec.~\ref{sec4}, we present an estimation of
the typical sample size of a $13^\circ$ hot spot as a function of
time. The estimate shown in Fig.~\ref{fig:poemma} gives an event rate
of $\Gamma\sim 250~{\rm yr}^{-1}$. A $13^\circ$ angular radius solid
angle covers a fraction $f_{\rm sky}\sim0.013$ of the sky. Within a
hot spot, one expects both background and source contributions, with a
ratio $f_{\rm events}=n_{\rm ev}/n_{\rm bg}$. With this, the required
life time of the experiment to measure a hot spot of $N$ events can be
roughly estimated to be \begin{equation} T\sim \frac{N}{\Gamma f_{\rm
      sky} f_{\rm events} }.\end{equation} For $f_{\rm events}\sim3,$
as observed in~\cite{Aab:2017njo} from the direction of Cen A,
$T\sim0.1 N\,{\rm yr}$.\footnote{We are not claiming that this is the
  value to expect, but just showing a possible value.} The projected
sensitivity of POEMMA is shown in Fig.~\ref{fig:powers}. For hot spots
of 20 or more events, the discovery power (with $\alpha=0.05$) is
almost one for nuclei other than helium. Therefore, we conclude that
if the hot spot is composed of nuclei heavier than nitrogen, in two
years of operation POEMMA will be able to exclude a pure-proton origin
at the 95\% CL.

\section{Conclusions}
\label{sec6}

In the spirit of~\cite{Anchordoqui:2017abg}, we have developed a
statistical test to quantify the ability of the future NASA's POEMMA
mission to isolate the nuclear composition of UHECRs using a subsample
of the distribution of arrival directions associated with a particular
source $\leftrightharpoons$ hot spot in the cosmic-ray-sky. This is possible
because sources of UHECR protons exhibit anisotropy patterns which
become denser and compressed with rising energy, whereas 
nucleus-emitting-sources give rise to a cepa stratis structure with
layers that become more distant from the source position with rising
energy. The peculiar shape of the hot spots from nucleus-accelerators
is steered by the competition between energy loss during propagation
and deflection on the GMF.

Our conclusions and caveats can be encapsulated as follows:
\begin{itemize}
\item We have shown that if an UHECR hot spot is composed of nuclei
  heavier than nitrogen, observation of roughly 20 events in this
  region of the sky will be required to discard a pure-proton
  explanation at the 95\% CL.
\item We have used the excess of events reported by the Auger
  Collaboration from the direction of Cen A~\cite{Aab:2017njo} to
  project that about 2~yr of POEMMA running will be necessary to probe
  the nuclear composition of this hot spot.
\item The magnetic field structure is not as simple as it has been
  considered here. It presents a highly anisotropic structure that would force us to consider its complexity in several ways: simulations should be performed individually for each source, propagating the cosmic rays from the source to Earth; and the distributions and data should include another angular variable to measure the orientation about the line of sight. 
\item The background should be considered in this picture. The
  presence of a background with or without a single nuclear composition would
  somewhat deteriorate our ability to reject a given hypothesis.
\item We have considered as known, and equal, the power law for all
  the energy spectra. Variations from this behavior could also have an
  effect on our results.
\item Both the angular and energy reconstruction resolution of POEMMA
  have to be considered in a complete analysis. The angular
  resolution, roughly estimated to be of the order of $1^\circ$,
  should not degrade the quality of our analysis significantly. The
  estimated $20\%$ energy resolution is also expected to have a minor
  impact.  Indeed the uncertainties introduced by these considerations
  would fall within errors of our working assumptions.
\end{itemize}

In summary, in a few years of operation the future NASA's POEMMA
mission will provide an {\it a priori} test of the evidence for hot
spots reported by the Auger~\cite{Aab:2018chp,Aab:2017njo} and TA~\cite{Abbasi:2007sv} collaborations. We have shown that
POEMMA satellite stereo observations will be able to determine
the UHECR composition using the distribution of arrival
directions. This new method to determine the nature of the particle
species is {\it complementary} to those using observables of extensive
air showers, and therefore is unaffected by the large systematic
uncertainties of hadronic interaction models.

\section*{Acknowledgments}
We thank our colleagues of the Pierre Auger and POEMMA collaborations
for some valuable discussions. The research of RCdA is supported by
the Fulbright Scholarship Program (Junior Faculty Member Award) and
CNPq under Grant 307750/2017-5. She also thanks the Physics and
Astronomy Department at Lehman College for hospitality. The research
of JFS and LAA is supported by the by the U.S. National Science
Foundation (NSF Grant PHY-1620661) and the National Aeronautics and
Space Administration (NASA Grant 80NSSC18K0464). TCP has also been
partially supported by NASA Grant 80NSSC18K0464. The research of DFT
is supported by grants AYA2015-71042-P, AYA2017-92402-EXP, iLink
2017-1238, and SGR 2017-1383. TADP and RJM were supported in part by
NSF grant AST-1153335. The research of FS and LW is supported by NASA APRA
NNX13AH55G. The POEMMA concept study is funded by NASA Award NNX17AJ82
at the University of Chicago and Goddard Space Flight Center. We
acknowledge the National Laboratory for Scientific Computing
(LNCC/MCTI, Brazil) for providing HPC resources of the SDumont
supercomputer, which have contributed to the research results reported
within this paper. URL: {\tt http://sdumont.lncc.br}. Any opinions,
findings, and conclusions or recommendations expressed in this
material are those of the authors and do not necessarily reflect the
views of the NSF or NASA.

\end{document}